\documentclass{WileyMSP-template}

\usepackage{nopageno}
\usepackage[utf8]{inputenc}
\usepackage[T1]{fontenc}
\usepackage[english]{babel}
\usepackage{hyperref}
\usepackage{tocloft}
\usepackage{graphicx}
\usepackage{caption,subcaption}
\usepackage{fancyhdr}
\usepackage{geometry,float}
\usepackage{mathtools,amsfonts,amsmath,amssymb}
\usepackage{braket}
\usepackage{siunitx}
\usepackage{lmodern}
\usepackage{multirow}
\usepackage{chngcntr}
\usepackage{setspace}
\usepackage{dsfont}
\usepackage{gensymb}
\usepackage{xcolor}
\usepackage{notes2bib}
\usepackage{underscore}
\usepackage{tabularx} 
\usepackage{comment}

\makeatother

\begin{document}

\pagestyle{fancy}
\rhead{\includegraphics[width=2.5cm]{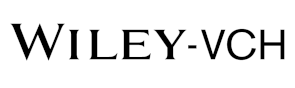}}

\title{Understanding Decoherence of the Boron Vacancy Center in Hexagonal Boron Nitride}

\maketitle

% Author: Please give full first and last names for authors and include * after the name of all corresponding authors

\author{András Tárkányi}
\author{Viktor Ivády*}

% Affiliations: Please provide adacemic titles (Prof. or Dr.) for all authors where applicable, and include an institutional email address for all corresponding authors
\begin{affiliations}
András Tárkányi, Asst. Prof. Viktor Ivády \\
\emph{Department of Physics of Complex Systems, E\"otv\"os Lor\'{a}nd University,\\
Egyetem t\'{e}r 1-3, H-1053 Budapest, Hungary } \\
András Tárkányi, Asst. Prof. Viktor Ivády \\
\emph{MTA–ELTE Lend\"{u}let "Momentum" NewQubit Research Group,\\
P\'{a}zm\'{a}ny P\'{e}ter, S\'{e}t\'{a}ny 1/A, 1117 Budapest, Hungary}\\

Email Address: ivady.viktor@ttk.elte.hu

\end{affiliations}

% Keywords: Please provide a minimum of three and a maximum of seven keywords, separated by commas


\justifying
% Abstract should be written in the present tense and impersonal style (i.e., avoid we), and be at most 200 words long
\begin{abstract}
\noindent

Hexagonal boron nitride (hBN) has emerged as a significant material for quantum sensing, particularly due to its ability to host spin active defects, such as the negatively charged boron vacancy (V$_\mathrm{B}^-$ center). The optical addressability of the V$_\mathrm{B}^-$ center and hBN's 2D structure enable high spatial resolution and integration into various platforms. 
%The V$\mathrm{B}^-$ center is sensitive to a range of parameters and operates at room temperature, making it an excellent candidate for quantum sensing.
However, decoherence due to the strong magnetic noise in hBN imposes fundamental limitations on the sensitivity of V$_\mathrm{B}^-$ center-based applications. Understanding the phenomena behind decoherence and identifying parameter settings that provide the highest performance are essential for advancing V$_\mathrm{B}^-$ sensors. This study employs state-of-the-art computational methods to investigate the decoherence of the V$_\mathrm{B}^-$ center in hexagonal boron nitride across a wide range of magnetic field values from 0~T up to 3~T. The provided in-depth numerical and analytical analysis reveals an intricate interplay of various decoherence mechanisms. This study identifies five distinct magnetic field regions governed by different types of magnetic interactions with and within the abundant nuclear spin bath. In addition to magnetic field, the effects of zero-field splitting, nuclear polarization, and different hyperfine coupling terms are studied, representing an important step forward in utilizing V$_\mathrm{B}^-$ ensembles in sensing. In particular, this study proposes operation in the moderate $180-350$~mT magnetic field range in chemically pure h$^{11}$B$^{15}$N samples, where the coherence time can reach $1-20$~$\mu$s, significantly exceeding the $\mathcal{O}( 100~\text{ns})$ low-field $T_2$ values.

\end{abstract}

% Text: Please use section headings and subheadings as specified below. For communications, all section headings apart from Experimental Section should be removed
% Please make the first reference to a display item bold: \textbf{Figure 1}
% Do not abbreviate Figure, Equation, etc.; display items are always singular, i.e., Figure 1 and 2.
% Equations are always singular, i.e., Equation 1 and 2, and should be inserted using the {equation} environment, not as graphics
% Please do not use footnotes in the text, additional information can be added to the Reference list.

\section{Introduction}

%\note{general questions: standing or italic T2, electron or electronic spin}  

%\V{use italic}
Recent advances in nanoscale sensing have enabled precise measurement of various physical quantities, such as magnetic field, electric field, temperature, strain, and pressure, at micrometer scales with high resolution.\cite{degen-quantum-2017,aslam-quantum-2023,du-single-molecule-2024} Point defect quantum sensors in semiconductors, in particular the negatively charged nitrogen-vacancy (NV) centers in diamond,\cite{doherty-nitrogen-vacancy-2013} leverage the quantum properties of localized defect states to achieve sensitivities and spatial resolutions beyond classical limits.\cite{degen-quantum-2017,schirhagl-nitrogen-vacancy-2014} NV-centered based sensors have demonstrated their capabilities in numerous applications in materials science and biology.\cite{schirhagl-nitrogen-vacancy-2014,dolde-electric-field-2011,boss-quantum-2017,rendler-optical-2017,hache-nanoscale-2025,petrini-nanodiamondquantum-2022}

The sensitivity of point defect quantum sensors is determined by the intrinsic coupling strength to the field to be measured, the coherence time of the electron spin, the sensor's layout, and characteristics of the readout mechanism.\cite{degen-quantum-2017} Bulk NV centers exhibit long coherence time, good signal to noise ratio due to their bright optical transition and sizable room temperature contrast, and considerable coupling strength of magnetic field, electric field, strain, and temperature changes. These advantageous properties, however, degrade in proximity to the diamond surface, due to charge instability and increased magnetic noise, which set the limits for sensitivity and spatial resolution in nanoscale detection.\cite{sangtawesin-origins-2019,bluvstein-identifying-2019,dwyer-probing-2022, pershin-shallowNV-2025} %In practical terms, the quality and distance from the surface play a crucial role in defining the capabilities of NV sensors. %On one hand, the diamond surface introduces significant electric and magnetic noise, which limits the lifetime and coherence time of quantum states when the NV center is close to the surface. \cite{dwyer-probing-2022} On the other hand, increasing the distance from the surface reduces the signal-to-noise ratio. \cite{du-single-molecule-2024} %The optimal depth for NV-based sensing is found to be approximately 13~nm\V{\cite{}}, which poses challenges for increasing spatial resolution. Surface roughness and uncontrolled defect orientation impose yet other difficulties utilizing nanodiamonds in sensing applications.\V{\cite{}}

In contrast, two-dimensional wide-bandgap semiconductors offer a smooth, low-noise surface while also allowing precise control over the sensor-target distance through tunable layer thicknesses.\cite{tetienne-quantum-2021,hassan-2d-2023,vaidya-quantum-2023} Moreover, van der Waals nature of 2D semiconductors enables integration into various platforms, including nanophotonic devices and layered heterostructures, offering a great host material for advanced sensing.\cite{tetienne-quantum-2021,gottscholl-spin-2021,healey-quantum-2022,kumar-magnetic-2022}

The leading contender in the field of defect-based van der Waals sensing is the negatively charged boron vacancy (V$_\mathrm{B}^-$) center in hexagonal boron nitride (hBN).\cite{gottscholl-initialization-2020,ivady-ab-2020} 
%hBN is a layered material composed of boron and nitrogen atoms arranged in a honeycomb lattice. Its ultra-wide 6~eV bandgap makes it an excellent host for color centers, such as the V$_\mathrm{B}^-$ center.
This spin-active defect exhibits a spin-triplet ground and excited state, with spin-dependent optical transitions that enable optical readout and manipulation.\cite{gottscholl-initialization-2020,ivady-ab-2020} V$_\mathrm{B}^-$-containing hBN flakes have demonstrated their ability to sense multiple physical parameters, such as temperature \cite{gottscholl-spin-2021}, pressure \cite{gottscholl-spin-2021}, strain \cite{lyu-strain-2022,yang-spin-2022}, electric field \cite{udvarhelyi-planar-2023}, and static magnetic fields \cite{gottscholl-spin-2021}, and to detect paramagnetic spins in liquids \cite{gao-quantum-2023}. Notably, these sensors can operate at room temperature \cite{gottscholl-spin-2021} in few-layer samples \cite{durand-optically-2023,robertson-detection-2023,clua-provost-impact-2024} and graphene-hBN-graphane van der Waals heterostructures \cite{fraunie-charge-2025}. These attributes make the V$_\mathrm{B}^-$ center in hBN a promising platform for advancing quantum sensing.

%V$_\mathrm{B}^-$ center is of particular interest for magnetic field imaging.
Recent reports have demonstrated wide field magnetic imaging of layered ferromagnets, such as CrTe$_2$ \cite{healey-quantum-2022,kumar-magnetic-2022} and Fe$_3$GeTe$_2$ \cite{huang-wide-2022}. In particular, Healey \emph{et~al.}~\cite{healey-quantum-2022} have demonstrated time-resolved imaging of temperature and magnetic field near the Curie temperature of CrTe$_2$ and mapped the charge currents and Joule heating in graphene. Exploiting spin relaxometry methods, Huang\emph{ et~al.}~\cite{huang-wide-2022} reported on spatially varying magnetic fluctuations in an exfoliated Fe$_3$GeTe$_2$ flake peaking around the Curie temperature in amplitude. Furthermore, Robertson \emph{et al.}~\cite{robertson-detection-2023} and Gao \emph{et al.}~\cite{gao-quantum-2023} have demonstrated detection of electron spin in liquid phase at room temperature through the quenching of the V$_\mathrm{B}^-$ center's spin relaxation time $T_1$. %Integration of V$_\mathrm{B}^-$ center-containing hBN flakes into 2D platforms facilitates the study of interfacial phenomena, such as spin transport in van der Waals heterostructures, and supports in vitro biological imaging of various physical properties {\cite{}}

%Despite these significant early results, the V$_\mathrm{B}^-$ center remains largely unexplored. For instance, its coherence properties have not been well characterized yet, however; the coherence time sets limits for the sensitivity of quantum sensors.\V{\cite{}} 
Despite these significant results, the coherent dynamics of the V$_\mathrm{B}^-$ center remain largely unexplained. 
%however; the coherence time sets limits for the sensitivity of quantum sensors.\V{\cite{}} \A{duplicate statement?}
%Because both boron and nitrogen belong to odd-numbered columns in the periodic table (III and V), all isotopes possess nonzero nuclear spins, creating a strongly coupled nuclear spin bath that can shorten coherence times and reduce sensing sensitivity. 
For instance, the reported Hahn-echo coherence times ($T_2$) range from tens of nanoseconds to tens of microseconds in different samples and conditions.\cite{VBmeas-liu2022,VBmeas-Gottscholl-2021,extending-Rizzato2023,isotope-Gong-2024,decohVB-Haykal-2022,cohprot-Ramsay2023,VBmeas3T-dyakonov-2022} While a shorter $T_2$ of 10–100~ns appears to be the intrinsic coherence time of the V$_\mathrm{B}^-$ center at low magnetic field, no study has systematically examined and explained the influence of relevant control parameters (such as magnetic field, strain, nuclear spin polarization) on extended scales and searched for decoherence protected settings exhibiting outstanding $T_2$ values. Addressing this knowledge gap in the literature is, however, not straightforward, since such an extensive experimental investigation is time and resource-consuming. Recent advances in cluster correlation expansion methods \cite{yang-quantum-2008,zhao-decoherence-2012,seo-quantum-2016,onizhuk-probing-2021} enabled reliable and efficient numerical simulation of the coherent dynamics of point defect spin qubits in semiconductors, which facilitates closing the knowledge gap in the literature.

In this work, we employ the recently developed generalized cluster correlation expansion (gCCE) method \cite{yang-quantum-2008,onizhuk-probing-2021,yang-longitudinal-2020} on convergent 1000 nuclear spin models to explore and understand the decoherence of the V$_\mathrm{B}^-$ center in h$^{11}$B$^{15}$N, which exhibits elongated coherence time and provide computational advances. To cover the full range of applicable magnetic fields, we compute coherence time from 0 to 3~T and reveal five different regions governed by different magnetic interactions serving as noise sources. %\V{We discuss the role of various isotopes and select the highest performing h$^{11}$B$^{15}$N host for our detailed study.}\A{we should rationalize the choice of isotopes} 
Our numerical results are supported by analytical arguments to provide a deep understanding of the dominant phenomena at the distinct regions. We also utilize analytical expressions to comprehend electron spin echo envelope modulation dictated by the three first-neighbor nitrogen nuclear spins and surrounding boron spins. In addition to the applied magnetic field, we investigate the effect of the transverse zero-field splitting parameter, polarization of the first-neighbor nuclear spins, and components of the hyperfine tensors on decoherence of the V$_\mathrm{B}^-$ center. Finally, relying on our numerical and analytical results, we discuss caveats in the literature and suggest exploiting the magnetic field region of $180-350$~mT where the coherence time of the V$_\mathrm{B}^-$ center in h$^{11}$B$^{15}$N reaches $1-20$~$\mu$s.

%\A{important things to mention: why Hahn echo, why these isotopes,} \A{should we explain Hahn-echo in this journal?} \V{Done, i guess.}

\section{Results and discussion}

In this article, we investigate the effect of nuclear spin-induced decoherence processes, which play a significant role due to the dense nuclear spin bath. Other paramagnetic defects in the proximity of the defect can further contribute to the magnetic noise and further reduce the coherence time. Therefore, our results can be considered theoretical upper bounds for the coherence time, representing the case of chemically and isotope-purified samples.

In the subsequent sections, we first review the approximation and first principles parameters obtained and used in this paper to compute numerical $T_2$ estimates. Next, we discuss our results and the underlying phenomena mostly for the case of h$^{11}$B$^{15}$N, which provides superior capabilities over other isotope compositions in many circumstances.

\begin{figure}[h!]
\begin{center}
\includegraphics[width=1.0\columnwidth]{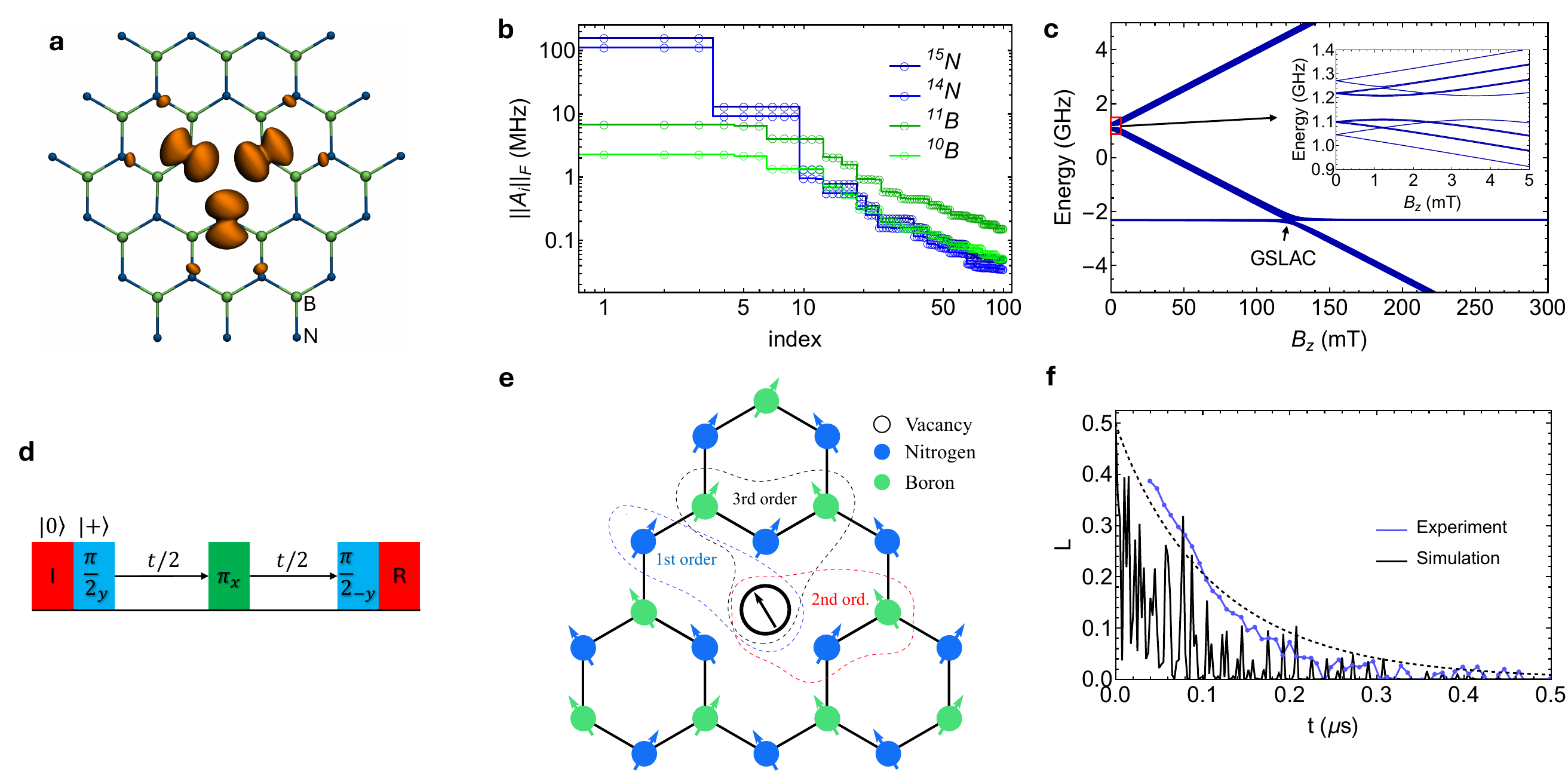}
\caption{ {\bf The V$_\mathrm{B}^-$ center in hexagonal boron nitride.} {\bf a} Structure and spin density of the V$_\mathrm{B}^-$ center in hexagonal boron nitride. Orange lobs depict the spin density distribution of the triplet electron spin localized on the dangling bonds of the neighboring nitrogen atoms. {\bf b} Frobenius-norm of the hyperfine coupling tensors of different species of nuclear spins in the spin bath. {\bf c} Ground state energy level structure of the V$_\mathrm{B}^-$ center electronic spin coupled to the three first-neighbor $^{15}$N nuclear spins. The inset shows the energy level structure close to zero-field. {\bf d} Schematic representation of the Hahn-echo pulse sequence used to measure decoherence due to dynamical magnetic noise. Red rectangles represent laser pulses, while blue and green rectangles represent microwave $\pi$-half and $\pi$ pulses, respectively. {\bf e} Illustration of first, second, and third-order clusters of the spin bath centered around the V$_\mathrm{B}^-$ electron spin. {\bf f} Comparison of measured and simulated coherence decay function at $B_z=0$~mT. Experimental data is reproduced from Ref.~\cite{VBmeas-liu2022} by rescaling the contrast values to the values of the coherence function $L$.}
\label{fig:fig1} 
\end{center}
\end{figure}

\subsection{Hyperfine interaction and cluster correlation expansion} 

The spin-1 electronic spin of the V$_\mathrm{B}^-$ center interacts with the surrounding nuclear spin bath through the hyperfine coupling, which plays a key role in our quantitative numerical analysis. The influence of the nuclear spin bath on electron spin dynamics is affected by the zero-field splitting and Zeeman interaction, which can either enhance or suppress certain sources of magnetic field noise depending on the splitting of the spin eigenstates. To establish the foundations for our discussion, we first provide accurate hyperfine coupling parameters and briefly review the fine and hyperfine energy level structure of the V$_\mathrm{B}^-$ center. 

Removal of a boron atom from the hBN 2D layer leaves three sp$^{2}$ like dangling bonds in the lattice. To understand the principles of the ground state V$_\mathrm{B}^-$ system, it is sufficient to consider these states only. In D$_{3\text{h}}$ point group symmetry, the three symetrically equivalent dangling bonds form an $a_1^{\prime} = (\sigma_1 + \sigma_2 + \sigma_3)/3$ state and a double degenerate $e^{\prime} = \left\lbrace (2\sigma_1 - \sigma_2 - \sigma_3)/4 \text{, } (\sigma_2 - \sigma_3)/2 \right\rbrace$ state. In the relevant negatively charged state of the V$_\mathrm{B}^-$ center, the $a_1^{\prime}$ state is fully occupied, while two electrons with aligned spins occupy the $e$ state. 
%The spin density, which corresponds to the electron density of the unpaired electrons, is primarily formed by the half-filled $e$ state, as illustrated in Figure~\ref{fig:fig1}a.
Density functional theory (DFT) \cite{ivady-ab-2020} calculations demonstrate that the spin density is mostly localized on the dangling bonds of the nitrogen atoms, see Figure~\ref{fig:fig1}a, although nonzero spin density can also be found at more distant neighboring lattice sites, e.g.\ on the six fourth-neighbor nitrogen atoms.

The hyperfine dipole-dipole coupling term can be obtained from integrating the electron spin density with the kernel of magnetic dipole-dipole interaction.
%kernel with a nuclear spin located at the lattice site $R_i$.\A{maybe a simpler description if this term is sufficient}
The Fermi contact interaction term, a second contribution to the hyperfine interaction, emerges when the electron spin density overlaps with the nucleus. Due to the spatial extension of the $e^{\prime}$ state, the latter term cannot be neglected. Recent developments have enabled the precise computation of both of these terms for an arbitrary number of lattice sites.\cite{hypfine-Takacs-2024} Building on these advancements, here, we calculate hyperfine tensors for 13000 $^{11}$B and $^{14}$N atomic sites (up to 3~nm distance from the center of the defect), which are available online through Ref.~\cite{Hyperfine-webpage}. For details of the DFT calculations, refer to the Methods section. From the comparison of the first neighbor nitrogen atoms' computed $A_{zz} = 47.14$~MHz hyperfine component with the experimental $A_{zz} = 47$~MHz \cite{gottscholl-initialization-2020} and $A_{zz} = 48$~MHz \cite{decohVB-Haykal-2022} values, we estimate the relative error in our hyperfine parameters smaller than 2\%, which is similar to the accuracy reported for the NV center in diamond \cite{hypfine-Takacs-2024}. We note that the hyperfine tensors of $^{15}$N nuclei can be straightforwardly obtained by scaling the tensors with the ratio of the nuclear g-factors. Such a high-precision hyperfine dataset is crucial for our quantitative numerical analysis and predictions. To gain insight into phenomena of decoherence, we occasionally employ the pseudo-secular approximation for the hyperfine tensor and keep only the vector $\vec{A}_{\text{ps}} = \left( A_{zx}, A_{zy}, A_{zz} \right)$.
%We note here that the point electron spin density approximation, often used to calculate coherence times, is inadequate for the V$_\mathrm{B}^-$ center due to the dominant role of short-range interaction in determining the $T_2$ coherence time, see numerical findings later. \A{unclear, to me at least}

In Figure~\ref{fig:fig1}b, we depict the Frobenius norm of the calculated hyperfine tensors $\mathbf{A}^{(i)}$ for the first hundred strongest coupled nuclear spins in decreasing order. Steps in the hyperfine Frobenius norm indicate \emph{hyperfine shells} (not necessarily coinciding with atomic shells of increasing distance), while the width of the steps indicates the multiplicity of the atomic sites within the shells. Considering the first three nitrogen and boron hyperfine shells, the Frobenius norm of the hyperfine tensors decreases by approximately two orders of magnitude. For the remaining shells, the hyperfine interaction is further reduced by $\sim$2~orders of magnitude. %The comparison of these results and the electronic-nuclear spin coupling within the NV center (3.1~MHz for $^{15}$N) clearly 
Since all lattice sites are occupied by paramagnetic isotopes, Figure~\ref{fig:fig1}b demonstrates the strong hyperfine coupling of the V$_\mathrm{B}^-$ to the nuclear spin bath. 

Characteristics of hyperfine interaction-related magnetic noise and the effect of single-spin or two-spin interactions are determined by the many-spin effective Hamiltonian
\begin{equation}
\begin{split}
    \hat{H}=&\hat{H}_e+\hat{H}_b + \hat{H}_{e-b} = D\left(\hat{S}_z^2-S(S+1)/3\right)+\frac{E}{2}\left(\hat{S}_+^2+\hat{S}_-^2\right)+g_e\mu_B\mathbf{B}^T\mathbf{\hat{S}}-\\
    &\sum_ig_N^{(i)}\mu_N^{(i)} \mathbf{B}^T\mathbf{\hat{I}}^{(i)}+\sum_i\mathbf{\hat{I}}^{(i)}\mathbf{Q}^{(i)}\mathbf{\hat{I}}^{(i)}+\sum_{i<j}\mathbf{\hat{I}}^{T(i)}\mathbf{J}^{(ij)}\mathbf{\hat{I}}^{(j)} + \sum_i \mathbf{\hat{S}}^T\mathbf{A}^{(i)}\mathbf{\hat{I}}^{(i)},
\end{split} \label{eq:hamil}
\end{equation}
where $\hat{H}_e$ and $\hat{H}_b$ are the electron spin and spin bath Hamiltonians, incorporating the electron spin zero-field splitting (ZFS) and Zeeman interaction terms, the nuclear spin Zeeman and quadrupole splitting, and nuclear spin-nuclear spin magnetic dipolar interaction terms, respectively. $\hat{H}_{e-b}$ is the interaction Hamiltonian describing hyperfine couplings between the V$_\mathrm{B}^-$ center and its environment. The parallel zero-field splitting parameter $D$ is set to 3470~MHz \cite{decohVB-Haykal-2022}, while the transverse ZFS parameter $E$ is set to the typical value 50~MHz \cite{gottscholl-initialization-2020}, unless explicitly stated otherwise in our study. 
%In \note{section ...} we show the dependence of coherence time on $E$ at low fields. 
For our calculations, the magnetic field vector is always set parallel to the c-axis of the defect, while for the quadruple splitting parameter of the $^{11}$B, we use the bulk value of $C_q = 3.72$~MHz obtained from our ab initio calculations. The nuclear dipole-dipole coupling tensor $\mathbf{J}^{(ij)}$ is calculated for all relevant nuclear spin pairs.

%In our numerical study, we focus predominantly on isotopically purified h$^{11}$B$^{15}$N, which is readily available on the market and has been studied experimentally\cite{}. We choose this composition as a primary target since the spin-3/2 $^{11}$B and the spin-1/2 $^{15}$N nuclei carry the smallest possible nuclear spin quantum number among the stable isotopes of boron and nitrogen. The resulting computational advantage facilitates the extensive numerical study carried out in this study. \V{On the other hand, we qualitatively discuss the effects of other isotope compositions at each magnetic field region in the upcoming sections.}

The fine structure of the spin energy levels of the V$_\mathrm{B}^-$ center as obtained from Equation~(\ref{eq:hamil})considering the electron spin and the closest three nitrogen $^{15}$N nuclear spins is depicted in Figure~\ref{fig:fig1}c. The energy levels fall into three branches according to the $m_S = \left\lbrace -1,0,1 \right\rbrace$ quantum number of the electron spin. There are two exceptions from the general rule; at zero magnetic field, the $m_S = \pm1$ states are mixed due to the transverse zero-field splitting parameter, therefore avoided crossings (clock transitions) emerge (see inset in Figure~\ref{fig:fig1}c) \cite{jamonneau-competition-2016, onizhuk-probing-2021,pershin-shallowNV-2025}. In the clock transition region, the defect spin is protected from magnetic field noises in first order. At the ground-state level anticrossing (GSLAC), the $m_S=-1$ and $m_S = 0$ states are mixed. In this study, we use the $m_S=-1$ and $m_S = 0$ braches as $\left | 0 \right \rangle $ and $\left | 1 \right \rangle $ qubit states, respectively. %At the avoided crossings, we use states of the smaller energy gap.

In our study, we compute the Hahn-echo coherence time $T_2$, which is the characteristic time scale of the intrinsic dynamical dephasing of the electron spin. The pulse sequence of the Hahn echo measurement is depicted in Figure~\ref{fig:fig1}d. First, a laser pulse initializes the electron spin in the $ \left| 0 \right\rangle$ state, which is then rotated into the $\left| + \right\rangle = \left| 0 \right\rangle + \left| -1 \right\rangle$ state with a microwave (MW) $\pi$-half pulse. The spin is let to evolve altogether $t$ time interrupted with a so-called refocusing $\pi$ pulse at halfway. The middle MW pulse is responsible for the cancellation of the phase shifts acquired from quasi-stationary magnetic noises. In an inhomogeneous ensemble (spatial, time, or both), the ensemble-averaged state vector refocuses into the pure $\left| - \right\rangle = \left| 0 \right\rangle + \left| -1 \right\rangle$ state after $t/2$ time of the $\pi$ pulse (spin echo). To measure the projection onto the $\left| - \right\rangle$ state, a final MW $\pi / 2$ pulse is applied to rotate the spin into the measurement basis, followed by a laser pulse converting the spin population information into photoluminescence intensity.
%\V{redundant...} When the magnetic noise is stationary, the Hahn echo pulse sequence eliminates the accumulated phase shifts in an ensemble. Therefore, the projection to the  $\left| - \right\rangle$ state is unity at time $t$ of the sequence. However, due to dynamical noises having an impact during the free evolution time $t$ of the electron spin, the projection of the refocused state onto the  pure $\left| - \right\rangle$ state falls below one. 
With increasing $t$, the dynamical noises have a larger impact resulting in a decaying echo signal characterized by decay time constant $T_2$, see for instance Figure~\ref{fig:fig1}f.  

Coherence properties of the electron spin can be theoretically obtained from the off-diagonal element of the reduced density matrix of the electron spin $\varrho_e (t) = \text{Tr}_{\text{bath}} \varrho (t)$ as 
\begin{equation}
    L(t) =  \text{Tr} \! \left( \sigma_+ \varrho_e (t) \right),
\end{equation}
where $\sigma_+$ is the raising operator. The complex-valued $L(t)$ function is the key quantity in our study. For visualization, we use the absolute value of $L(t)$. Due to the large number of nuclear spins involved, the exact time evolution of the system's density matrix is intractable, approximations need to be used. For the calculation of coherence properties, the state-of-the-art approach is the cluster correlation expansion (CCE) method \cite{yang-quantum-2008,zhao-decoherence-2012} that has proven its capabilities for several point defect spin quantum bits \cite{decohVB-Haykal-2022,seo-quantum-2016,onizhuk-probing-2021}. Here, we use our in-house implementation of the generalized cluster-correlation expansion (gCCE) method \cite{yang-quantum-2008,onizhuk-probing-2021,yang-longitudinal-2020}, which further extends the capabilities of the original CCE method. In cluster-correlation expansion, the coherence function $L(t)$ is expressed as a product of $L_n(t)$ coherence functions of different orders as
\begin{equation} \label{eq:Lprod}
    L(t) = \prod_{n} L_n(t). 
\end{equation}
To define and calculate $L_n(t)$, the many-spin system is divided into small clusters consisting of the electron spin (in all cases) and a few nuclear spins. An order $n$ cluster includes $n$ nuclear spins as illustrated in Figure~\ref{fig:fig1}e. In CCE, $L_n(t)$ accounts for the irreducible contribution of all possible $n^{\text{th}}$ order clusters to the coherence function and is expressed as  
\begin{equation}
    L_n(t) = \prod_{j\in C_n} \tilde{l}_j^{(n)}(t), 
\end{equation}
where $C_n$ defines the set of order $n$ clusters and $\tilde{l}_j^{(n)}(t)$ is the irreducible contribution of a single cluster defined as
\begin{equation}
    \tilde{l}_j^{(n)} = \frac{\text{Tr} \! \left( \sigma_+ \text{Tr}_{\text{bath}} \! \left( \varrho_j^{(n)} \right) \right)}{l^{(0)} \prod_{p\in C_{1,j}} \tilde{l}_p^{(1)} \cdots \prod_{q\in C_{n-1,j}} \tilde{l}_q^{(n-1)}}, 
\end{equation}
where $C_{m,j}$ defines the subset of order $m$ clusters that are contained within cluster $j$ of order $n$ ($0<m<n$), and $l^{(0)}$ is the conference function of the isolated electron spin. The normalization ensured that an order $n$ irreducible contribution solely describes $n$ particle correlation effects. Accordingly, the zero-order coherence function $L_0(t) = l^{(0)}(t)$ describes the Larmor precession of the electron spin ($|L_0(t)| = 0.5$ in the Hahn echo sequence), the first-order coherence function $L_1(t)$ accounts for decoherence effects derived from electron spin-nuclear spin coupling and correlation, the second-order coherence function $L_2(t)$ accounts for decoherence effects derived from nuclear spin-nuclear spin coupling and three-spin correlated magnetic noise, while third and higher order conference functions account for magnetic noise arising from four and more spin correlations.  

The CCE ansatz is exact in $N^{\text{th}}$ order for an $N$ nuclear spin central spin system. Approximations are introduced in practice through the truncation of the product in Equation~(\ref{eq:Lprod}) and through the neglect of the clusters containing distinct spins. Indeed, when higher-order magnetic noises have negligible effects within the coherence time of the system, the corresponding coherence functions $L_n(t) \approx 1$ for most $t$ values. The approximation order and the numerosity of the cluster in each order need to be checked rigorously for high numerical accuracy. In our numerical simulations, we check at all relevant magnetic field regions the convergence of approximation order and the relevance of the distinct nuclear spins in the cluster expansion, see Supporting Information for more details.

\subsection{Magnetic field dependence of coherence time: an overview} 

\begin{figure}[h!]
\begin{center}
\includegraphics[width=\columnwidth]{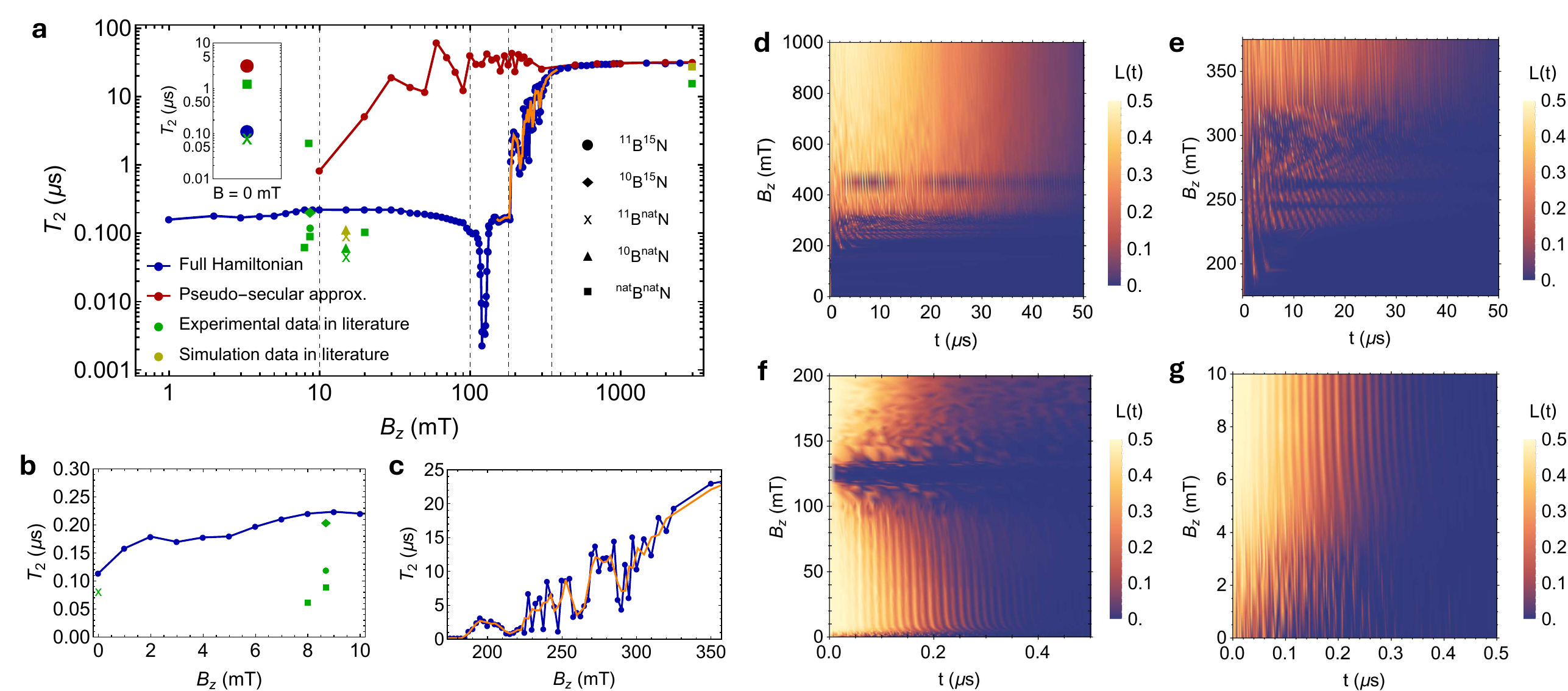}
\caption{{\bf Magnetic field dependence of the Hahn-echo coherence time ($T_2$) as a function of the external magnetic field.} {\bf a} Calculated coherence time from 1~mT to 3000~mT on log-log scale (blue). The orange curve serves as a guide to the eye. The results obtained in the pseudo-secular approximation of the hyperfine couplings are shown in red. Dashed lines indicate approximate boundaries of different regions dominated by different decoherence mechanisms. Green and yellow markers indicate experimental and simulation data available in the literature (see also Table \ref{tab:prev}). The inset shows the coherence times at zero-field. {\bf b, c} $T_2(B_z$) on a linear scale close to zero-field and in the transition region. {\bf d, e, f, g} Density plot of the calculated coherence function $L$ as a function of the spin echo time $t$ and the external magnetic field $B_z$ in different regions and on relevant timescales. Gold regions are indicators of coherent V$_\mathrm{B}^-$ centers, while dark blue regions indicate the loss of coherence. Transition and periodic modulation of the coloring function represent decay and electron spin echo envelop modulation, respectively.
\label{fig:fig2} }
\end{center}
\end{figure}

\renewcommand{\arraystretch}{1.5}

\begin{table}[h!]
    \centering
    \begin{tabular}{cc|c|c|c}
    \hline \hline
      & & Experiment& Simulation & This work \\
       \hline
       $B_z$  & Composition & $T_2$   & $T_2$ & $T_2$  in h$^{11}$B$^{15}$N \\
       \hline
    
       0 mT  & $^{11}$B$^{\text{nat}}$N & 82 ns$^{a}$ & &  \multirow{2}{*}{114 ns}\\
       0 mT  & $^{\text{nat}}$B$^{\text{nat}}$N & 1.2  $\mu$s$^{b}$ &  & \\
        \hline
       8 mT & $^{\text{nat}}$B$^{\text{nat}}$N & 60 ns$^{c}$ & &  220 ns \\
       8.5 mT & $^{\text{nat}}$B$^{\text{nat}}$N & 2 $\mu$s$^{b}$ & &  222 ns \\
       8.7 mT & $^{\text{nat}}$B$^{\text{nat}}$N &  87 ns$^{d}$ & &  \multirow{3}{*}{223 ns} \\
       8.7 mT & $^{11}$B$^{15}$N & 119 ns$^{d}$ &  &\\
       8.7 mT & $^{10}$B$^{15}$N & 205 ns$^{d}$ &  &\\
        \hline
       15 mT  & $^{11}$B$^{\text{nat}}$N & 46 ns$^{c}$ & 92 ns$^{e}$ &  \multirow{2}{*}{220 ns} \\
       15 mT & $^{10}$B$^{\text{nat}}$N & 62 ns$^{c}$ & 115 ns$^{e}$ &  \\
       20 mT & $^{\text{nat}}$B$^{\text{nat}}$N & 100 ns$^{f}$ & &  222 ns\\
        \hline
       3 T & $^{\text{nat}}$B$^{\text{nat}}$N & 15 $\mu$s$^{g}$ & 27 $\mu$s$^{h}$ &  31 $\mu$s\\
       \hline\hline
    \end{tabular}
    \caption{{\bf Measured and calculated Hahn-echo coherence times available in the literature.} $B_z$ is the applied magnetic field along the quantization axis, "composition" refers to the isotopic composition of the sample, and $T_2$ is the Hahn-echo coherence time. References: $^{a}$Ref.~\cite{VBmeas-liu2022}, $^{b}$Ref.~\cite{VBmeas-Gottscholl-2021}, $^{c}$Ref.~\cite{extending-Rizzato2023}, $^{d}$Ref.~\cite{isotope-Gong-2024} , $^{e}$Ref.~\cite{decohVB-Haykal-2022}, $^{f}$Ref.~\cite{cohprot-Ramsay2023}, $^{g}$Ref.~\cite{VBmeas3T-dyakonov-2022}, $^{h}$Ref.~\cite{fpdecoh-Lee-2022}.
    }
    \label{tab:prev}
\end{table}

To explore the dependence on an external magnetic field applied along the quantization axis of the V$_\mathrm{B}^-$ center, we calculate Hahn-echo coherence times from $B_z=0$~T to $B_z=3$~T, see Figure \ref{fig:fig2}a. After analyzing the coherence plot, we define five regions showing qualitatively different coherence dynamics. \emph{Close to zero-field},
%assuming a \V{50~MHz} transverse zero-field splitting, 
$T_2$ increases monotonically with increasing magnetic field strength up to $\sim$10~mT, where it reaches a plateau of $T_2 \approx 200$~ns, see Figs~\ref{fig:fig2}b and g. In the \emph{low-field regime} (between 10~mT and 100~mT), $T_2$ stays constant for a wide region before it starts to decrease when approaching the GSLAC. In this interval, the coherence function $L(t)$ is a Gaussian modulated by a magnetic field-independent oscillation of a frequency $\sim$67~MHz, see Figure~\ref{fig:fig2}f. By comparing the coherence times obtained in the pseudo-secular approximation of the hyperfine coupling (red curve in Figure~\ref{fig:fig1}a) with the full gCCE solution, we observe a large deviation showcasing the dominant rule of electron spin-flipping terms of the hyperfine interaction in the decoherence processes. We demonstrate later that nuclear spin precession and electron spin-mediated nuclear spin flip-flops govern the decoherence in this region. In the third \emph{region around the GSLAC}, the coherence time decreases by more than an order of magnitude, which is a consequence of the strong mixing of the electron and nuclear spin states and the resulting rapid relaxation of the qubit states. By neglecting the electron spin-flipping terms of the hyperfine interaction, the drop in coherence time at the avoided crossing is not observed, see red curve in Figure~\ref{fig:fig2}a. We note that the introduction of the pseudo-secular approximation of the hyperfine interaction yields essentially the CCE method, which, in contrast to gCCE, ignores electron spin flipping processes. Thus, our results for magnetic fields below 350 mT demonstrate that the gCCE method needs to be employed to capture the coherence time. We also note that while the gCCE method qualitatively captures the experimentally observable phenomenon at the GSLAC, it underestimates coherence times in this region \cite{coherence-avoided-cr-Onizhuk-2021}.

Strikingly, right after the GSLAC at $B \approx 180$~mT, we observe \emph{the transition region}, where the coherence time increases rapidly and, for a moderate magnetic field value of $B_z = 350$~mT, it reaches $T_2 \approx 23$~$\mu$s, which is two orders of magnitude larger than the largest $T_2$ value at low magnetic fields, see Figure~\ref{fig:fig2}a, c, and e. The increase of the coherence time is, however, nonmonotonous and the coherence functions are highly irregular at the transition region from low-field to high-field regime. After a rapid Gaussian decoherence, we observe a partial recovery of the coherence function exhibiting slow decay, see Figure~\ref{fig:fig2}e. Similar behavior has been observed for the silicon vacancy in silicon carbide \cite{4hsic-Yang-2014}. In the \emph{high-field regime}, where $B_z\gtrsim350$~mT, the decay of $L$(t) is again a Gaussian with negligible envelope modulation. Both the gCCE and the pseudo-secular coherence times saturate at $T_2\approx31~\mu$s, indicating that hyperfine-induced spin flipping does not play a role in this region.

Next, we compare our numerical results with experimental and previous theoretical results available in the literature. We find limited experimental reports on the magnetic field dependence of the $T_2$ time, especially for the V$_\mathrm{B}^-$ center in h$^{11}$B$^{15}$N considered in our study. Table~\ref{tab:prev} summarizes the literature on $T_2$ measurement and simulation in various samples. As can be seen, our results compare well with recently published experimental and simulated values; however, in almost all cases, the values obtained in our simulations are superior to the literature, due to the isotope composition used and the neglect of electron spin-related magnetic noise. 
%Table~\ref{tab:prev} also highlights the important role that isotopic composition plays in coherence dynamics, in line with previous studies \cite{decohVB-Haykal-2022, isotope-Gong-2024, fpdecoh-Lee-2022}. 
%The reason for the slight overestimation observed in our study is twofold.
By comparing to our previous numerical results \cite{decohVB-Haykal-2022} on the V$_\mathrm{B}^-$ center in h$^{11}$B$^{14}$N and h$^{10}$B$^{14}$N at 15~mT, we see that the use of $^{15}$N nuclear spin enhances coherence time by a factor of $\sim$2 at this magnetic field. A better coherence time in $^{15}$N samples seems to be confirmed by experiments at 8.7~mT, where longer coherence time is reported in h$^{11}$B$^{15}$N than in h$^{\text{nat}}$B$^{\text{nat}}$N. We recall here that natural hBN contains 80.1\% $^{11}$B and 99.6\% $^{14}$N nuclear spin, thus the main difference in the nuclear spin bath in these measurements is the nitrogen isotope content. %The second reason for the overestimation when comparing our results with experimental data is that our spin bath model takes into account boron and nitrogen nuclear spins only and neglects the effect of different paramagnetic impurities, electronic spins, and other contaminants present in the host material. 
Due to the neglect of electron spin noises, our numerical results can be considered as theoretical upper bonds for the coherence time available in chemically pure h$^{11}$B$^{15}$N samples.

Finally, we draw attention to the very first $T_2$ measurement carried out on the V$_\mathrm{B}^-$ center in Ref.~\cite{VBmeas-Gottscholl-2021} that yielded long, 1.2~$\mu$s and 2~$\mu$s coherence times at 0~mT and 8.5~mT. Our theoretical results and other experiments in agreement provide an order-of-magnitude shorter coherence time. 
%After finishing our detailed analysis of the root cause of the decoherence of the V$_\mathrm{B}^-$ electron spin,
We provide possible explanations for these outlying experiments in the Conclusion. 

In the following sections, we thoroughly analyze the dominant spin interactions and noise sources in the above-defined regions and study parameter dependence beyond the magnetic field.

%For the fields $B_z<B_{GSLAC}$, despite the different isotopic compositions ($^{10}$B$^{15}$N, $^{11}$B$^{14}$N) from the system I studied ($^{11}$B$^{15}$N), the measured and simulated values are nearly identical, so the effect of the specific type of isotope is not significant in this range. In contrast, the measurement on the $3$ Tesla, $^{\mathrm{nat}}$B$^{14}$N sample is about half of the simulated value. The reason for this is the increased number of coherence loss channels due to the higher spin quantum number of $^{10}$B and $^{14}$N isotopes, see Figure \ref{sec:highfield} section.

\subsection{Decoherence close to zero-field}

%\V{Ezt a címet én adtam? Szerintem nem néz ki jól / nem elfogatott hogy a címben zárójel van.}

\begin{figure}[h!]
    \centering
    \includegraphics[width=\linewidth]{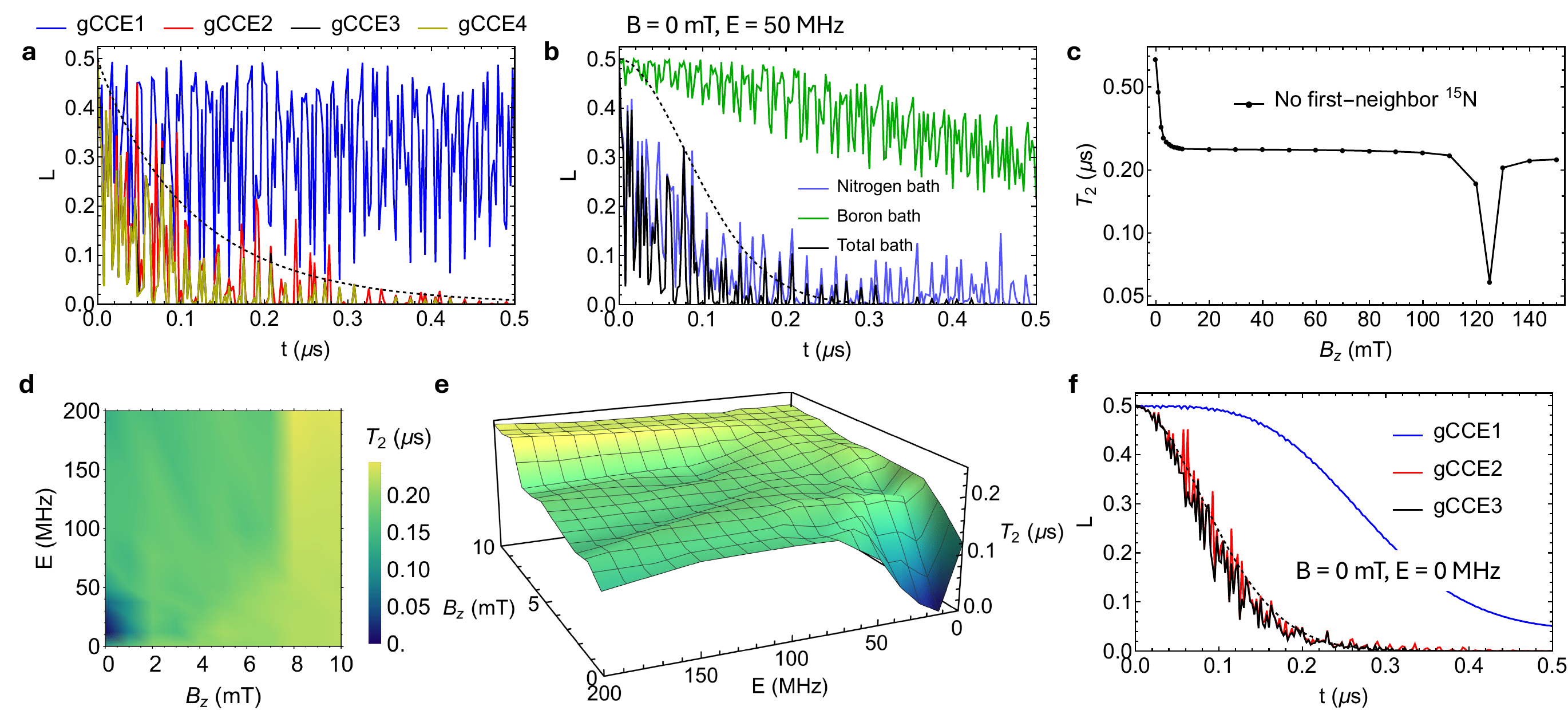}
    \caption{{\bf Hahn-echo coherence time close to zero-field for different configurations and interaction parameters.} {\bf a} Coherence functions in different orders of gCCE at zero-field. %and for measured transverse zero-field splitting.
    {\bf b} Coherence functions obtained by considering pure nitrogen ($^{15}$N), pure boron ($^{11}$B), and the combined boron-nitrogen spin bath. {\bf a} and {\bf b} The dashed line shows the fitted decay curve to the convergent $L(t)$ curve. {\bf c} $T_2$ as a function of magnetic field in a hypothetical system, where first-neighbor nitrogen spins are removed. {\bf d, e} Coherence time as a function of the transverse zero-field splitting ($E$) and the external magnetic field. {\bf f} The coherence function in different orders of gCCE at zero-field in the case of ideal threefold rotational symmetry of the defect.}
    %$T_2$ against the initial, uniform population of $\ket{\downarrow}$ nitrogen nuclear spin states.
    \label{fig:fig3}
\end{figure}

First, we discuss the coherence dynamics observed close to zero-field, namely within the $B_z=0-10$~mT magnetic field interval. We examine the contribution of the different gCCE orders to the coherence function at zero magnetic field, see Figure~\ref{fig:fig3}a. The noninteracting bath approximation (gCCE1) yields a not-decaying, oscillating coherence function $L_1 (t)$, while the noises emerging from second- (gCCE2) and third-order (gCCE3) clusters cause rapid decoherence, see Figure~\ref{fig:fig3}a. Irreducible correlations involving four nuclear spins (gCCE4) are not relevant factors in the decoherence at zero-field.

To study the effects of different nuclear spin species, we investigate decoherence in hypothetical systems, where either boron or nitrogen nuclei are removed from the bath, forming a pure $^{15}$N or $^{11}$B spin bath, respectively. Our results show that boron-induced decay of the coherence function is much slower compared to the nitrogen spin bath induced decay, and thus interactions with nitrogen spin pairs and triplets serve as the main source of decoherence, see Figure~\ref {fig:fig3}b. We note that close to zero-field, boron-nitrogen interactions also contribute to decoherence, see Supporting Information.

Interestingly, the expected increase of the coherence time at the coherence-protected clock transition region at $B \approx 0$~mT is not observed for the V$_\mathrm{B}^-$ center. To locate the source of this phenomenon, we calculate $T_2$ in the hypothetical system, where the three first-neighbor nitrogen spins are removed from the spin bath, see Figure~\ref{fig:fig3}c. In this case, the coherence time increases to $T_2\approx0.7$~$\mu$s at $B_z=0$~mT. Based on these observations and comparison with the results obtained in the pseudo-secular approximation, at zero magnetic field we derive three- and four-spin correlations as the dominant noise sources. These processes are induced by electron-nuclear state mixing due to nonsecular components of the hyperfine interaction with first-neighbor (and further) nitrogen spins, which we identify as primary (and secondary) sources of decoherence.

\subsubsection{Effect of the transverse zero-field splitting}

To further comprehend the absence of the coherence-protected region at $B_z = 0$~mT, we investigate the dependence of $T_2$ on the magnitude of the transverse zero-field splitting $E$, see Figure~\ref{fig:fig3}d and e. For increasing transverse zero-field splitting at $B_z = 0$~mT, we first observe a sharp drop in the coherence time. This trend is reversed at $ E \approx 12.5$~MHz, from which point further increase of the $E$ splitting elongates the coherence time. As the $E$ value exceeds the order of $\sim100$ MHz of the nonsecular hyperfine coupling of the first-neighbor nitrogens, the coherence time saturates to $T_2\approx0.2~\mu$s, limited by the fluctuating magnetic noise of boron spins, see Section \ref{sec:lowfield}.

At zero magnetic field and $E$ splitting, see Figure~\ref{fig:fig3}f, we obtain coherence curves that largely deviate from those obtained for a finite E splitting, see Figure~\ref{fig:fig3}a. Due to the disappearing clock transitions, the first-order magnetic noises are not suppressed for $E = 0$~MHz, thus, the gCCE1 coherence function exhibits a decaying function with no envelope modulation. At the same time, the higher-order gCCE curves exhibit moderate modulation of a Gaussian envelope, which is in contrast to the highly modulated exponential decay curve for $E = 50$~MHz.

By analyzing the observations, we conclude that the clock transitions do in fact protect the defect spin from the magnetic noise of single nuclear spins. However, higher order correlations cause significant decoherence through $\ket{-1}\leftrightarrow\ket{+1}$ double defect spin quantum jumps enabled by the $\hat{S}_+^2$, $\hat{S}_-^2$ operators. These transitions are primarily driven by effective interactions arising from the coupling of the zero-field splitting Hamiltonian and the nonsecular hyperfine interaction with two nuclear spins. For $E=0$ MHz, the double spin-flipping operators do not appear in the Hamiltonian, thus, the $\Delta m_S=\pm2$ transitions are forbidden in first order. This leads to an increased $T_2\approx0.2$~$\mu$s with coherence functions exhibiting behavior analogous to those obtained in the low-field regime, see Figure~\ref{fig:fig3}f. This leads us to the conclusion that, 
%besides the fluctuating magnetic noise of boron spins, 
central spin-mediated interaction between nuclear spin pairs becomes an additional source of decoherence (refer to Section \ref{sec:lowfield} for further discussion). For sufficiently large $E$ ($B_z$), the $\ket{-1}\leftrightarrow\ket{+1}$ transitions become suppressed within the timescales of other decoherence processes due to the opening of energy gaps surpassing higher order, hyperfine-driven mixing effects.

\subsection{Low-field regime}\label{sec:lowfield}
%\A{oscillation in Wei Liu?} \V{What do you mean?}

\begin{figure}[h!]
    \centering
    \includegraphics[width=\linewidth]{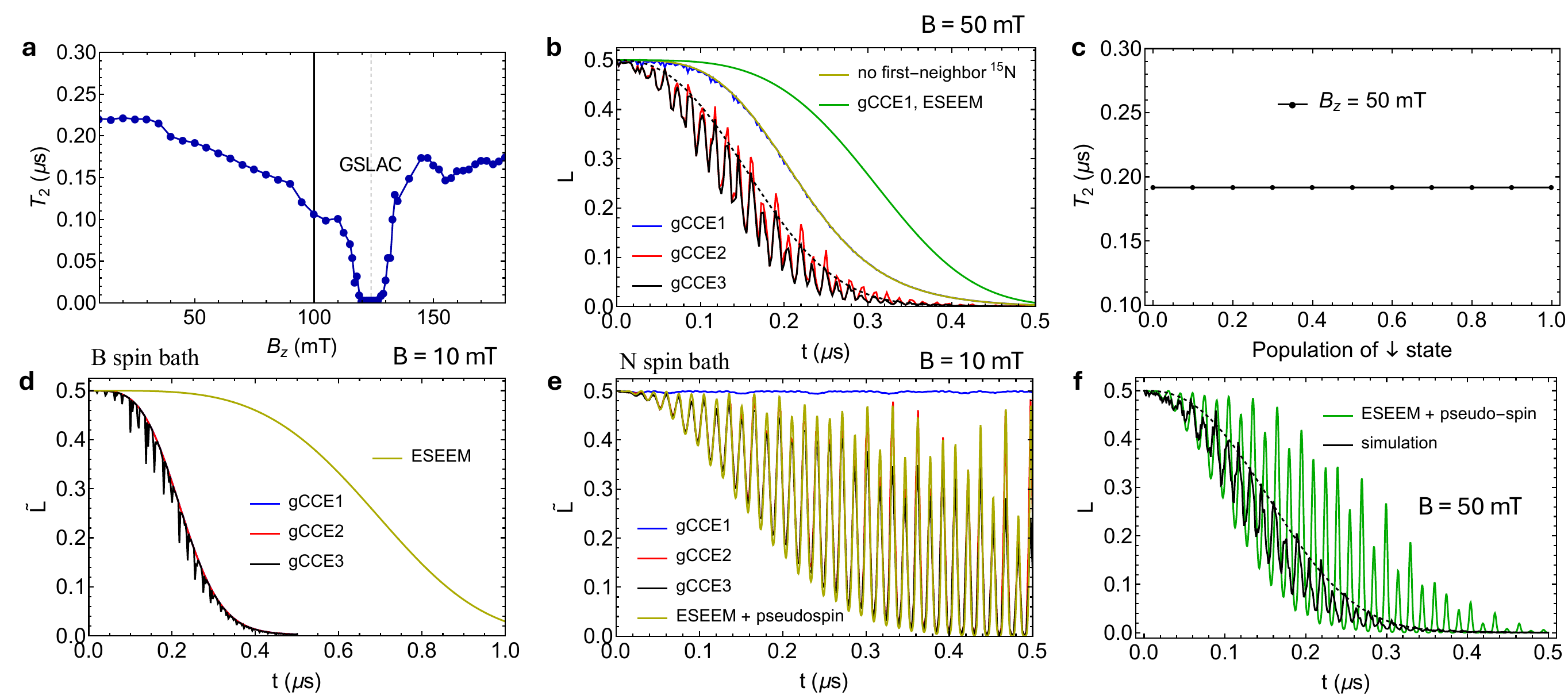}
    \caption{{\bf Analysis of coherence decay in the low-field regime.} {\bf a} Coherence time as a function of magnetic field in the low-field regime. {\bf b} Coherence functions in different orders of gCCE, the coherence function obtained by neglecting the role of all three first-neighbor nitrogen spins, and the first-order coherence decay derived from ESEEM theory. The dashed line shows the fitted decay. {\bf c} $T_2$ against the initial population of $\ket{\downarrow}$ states of first-neighbor nitrogen nuclear spins. {\bf d, e} Coherence functions in different orders of gCCE and analytical results obtained from ESEEM theory for spin baths consisting of solely boron and nitrogen nuclei, respectively. {\bf f} Coherence function obtained from simulations compared to the result of ESEEM and pseudo-spin theory.} 
    %Second- and third-order decoherence contribution considering the total Hamiltonian and in the pseudo-secular approximation.
    %\V{Move these to the SI.} {\bf g, h, i}  Comparison between second-order decoherence contributions obtained from simulations and derived considering effective interactions and pseudo-spin theory at several magnetic field values within the low-field regime indicated in the top-right corner of each subfigure.}
    \label{fig:fig4}
\end{figure}

After plateuing at a local maximum of $T_2 = 220$~ns in the range $B_z = 8-35$~mT, the Hahn-echo coherence time is only slightly affected by the magnetic field, and remains in the range of $T_2\approx0.1-0.2~\mu$s in the full interval (from 10~mT up to 100~mT), see Figure~\ref{fig:fig4}a. First-order clusters containing the defect spin and one nuclear spin induce a smooth Gaussian coherence function, while second-order clusters further reduce the coherence time and introduce a modulation of the coherence function with a frequency independent of the magnetic field applied, see Figure~\ref{fig:fig4}b and \ref{fig:fig2}f. The contribution from third-order clusters is negligible in the first half of the region, and slowly increases as the external field approaches the GSLAC.

As observed for other diatomic host materials \cite{seo-quantum-2016}, for magnetic fields larger than $\sim$10~mT, the boron and nitrogen spin bathes are decoupled due to the difference in the nuclear g-factors and the increasing difference of the nuclear Zeeman splitting, see Supporting Information for further details. Interestingly, in the low-field regime, the first- and second-order contributions are predominantly the consequence of noise originating solely from boron and nitrogen spins, respectively. This effect is demonstrated in Figure~\ref{fig:fig4}d and e. As can be seen, the spin bath-separated coherence function $L_{\text{N}}(t)$, obtained by considering the nitrogen spins only, exhibits no decay in first-order gCCE, meaning that electron spin induced nuclear spin noise, i.e.\ nitrogen nuclear spin precession has negligible effects. The second and third-order gCCE coherence funtions exhibit slow decay with significant envelope modulation connected to strong hyperfine interactions. In contrast, the boron-only coherence function $L_{\text{B}}(t)$ decays in the first-order, while exhibiting small modulations at higher orders. This indicates that the dominant noise source comes from the electron spin-induced precession of boron nuclear spins, and the negligible envelope modulation indicates the absence of strongly coupled boron. Consequently, in the $B_z=10-100$~mT magnetic field interval, the decoherence of the V$_\mathrm{B}^-$ center is dominated by the boron nuclear spins, while the nitrogen nuclear spins induce envelope modulation.

In the following sections, we carry out electron spin-echo envelope modulation (ESEEM) analysis for both the boron and nitrogen spin bath and derive second-order effective interactions to further comprehend and support our numerical results. 

\subsubsection{ESEEM theory analysis of coherence functions in the low-field regime}

Before discussing the ESEEM theory of the boron and nitrogen nuclear spin bath, three additional properties of the system need to be taken into consideration. First, due to symmetry reasons, the $A_{xz}$, $A_{yz}$, $A_{zx}$, and $A_{zy}$ components of the hyperfine tensors of in-plane nuclei (with respect to the defect) are zeros \cite{seo-first-principles-2024}. These terms induce an effective field tilted away from the quantization axis, which is then responsible for the electron and nuclear spin precession driven by the other spin. Consequently, neither the electron nor the nuclear spins precess for in-plane arrangements. In gCCE, this is signaled by the absence of the decay in the first-order. A decay in this order is induced by nuclear spins located in the planes above and below the plane containing the defect. Second, the g-factor of the $^{11}$B nuclear spin is $\sim$3.2 times larger than that of the $^{15}$N nuclear spin. In addition, the $^{11}$B spin itself is also 3 times larger than the doublet $^{15}$N nuclear spin. For the dipole-dipole hyperfine interaction term dominating at larger distances, this translates into an order of magnitude difference in the strength of off-diagonal elements of the hyperfine coupling Hamiltonian. Third, due to its larger electron affinity, the spin density is predominantly localized on nitrogen atoms in the plane of the vacancy, thus the isotropic Fermi contact hyperfine contribution is only significant for the nitrogen nuclear spin in close proximity to the defect in the plane. 

%\vspace{0.5cm}

%\noindent{\emph{Fluctuating magnetic noise of out-of-plane boron nuclear spins.}}

The first-order expansion in gCCE accounts for hyperfine interactions only and neglects defect spin-mediated correlations or couplings within the spin bath. The coherence function in such cases can be expressed in closed form using electron spin-echo envelope modulation theory. For a spin-$I$ nuclear spin bath and for a pure-dephasing (i.e.\ pseudo-secular) Hamiltonian the Hahn-echo ESEEM formula can be written in the approximate form of \cite{quadeseem-SHUBIN-1983}
\begin{equation}
    L_1(t)=\prod_i\left(1-\frac{8}{3}I_i(I_i+1)k_i^2\mathrm{sin}^2\left(\sqrt{(\omega_i+A_{\parallel,i})^2+A_{\perp,i}^2}\frac{t}{4}\right)\mathrm{sin}^2\left(\omega_i\frac{t}{4}\right)\right),
    \label{eq:ESEEM1}  %\label{eq:beating}
\end{equation}
where $A_\parallel=A_{zz}$, $A_\perp =\sqrt{A_{zx}^2+A_{zy}^2}$, $k^2=A_\perp ^2/((\omega_i+A_\parallel)^2+A_\perp^2)$ 
%are quantities related to the hyperfine coupling of the nuclear spins,
and $\omega=\gamma B_z$ is the Larmor frequency of the nuclear spin \cite{sic-divacancy-Seo}. According to Equation~(\ref{eq:ESEEM1}), the first-order coherence decay is a result of the superposed modulation caused by the fluctuating magnetic noise of precessing nuclear spins, with a frequency spread determined by the distribution of pseudo-secular hyperfine coupling terms ($A_{zx}$, $A_{zy}$). As discussed above, $A_\perp $ vanishes for the spins in the same layer as the vacancy, therefore, $k_i = 0$ and these nuclear spins do not modulate the coherence function in first-order. This explains our results in Figure~\ref{fig:fig4}e, showing no visible decay for $^{15}$N spin bath in gCCE1. In contrast, for the boron spin bath that couples mainly through the hyperfine dipolar interaction, an order of magnitude stronger than nitrogen at similar distances, we expect a decay with similar characteristics given by Equation~(\ref{eq:ESEEM1}). Indeed, both ESEEM theory and gCCE1 simulation give rise to stretched Gaussian decay, however, $T_2$ obtained by the former is $\sim4$ times longer, see Figure~\ref{fig:fig4}d. We attribute this discrepancy to the inexact treatment of the nonzero quadrupole splitting in Equation~(\ref{eq:ESEEM1}). The additional splittings in the energy level structure of boron nuclear spins create further decoherence channels at low fields. This effect decreases with increasing magnetic field, see Figure \ref{fig:fig4}b.
%The formula also explains the equality between $L$(t) obtained by completely neglecting the effect of the three first-neighbor nitrogen spins in ESEEM theory and in gCCE1, see Figure \ref{fig:fig4}b. 

%The gyromagnetic ratio of $^{15}$N is an order of magnitude smaller than of $^{11}$B, see Supporting Information, thus, owing to the ($\mathrm{sin}^2$) factor, the modulation caused by these spins is two orders of magnitude smaller.

%For $I>1/2$ nuclear spins the
%\begin{equation}
%    L_1(t)=\prod_i\left(1-\frac{8}{3}I_i(I_i+1)k_i^2\mathrm{sin}^2\left(\sqrt{(\omega_i+A_i)^2+B_i^2}\frac{t}{4}\right)\mathrm{sin}^2\left(\omega_i\frac{t}{4}\right)\right)
%    \label{eq:beating}
%\end{equation}
%approximate formula can be applied \cite{quadeseem-SHUBIN-1983}. The resulting theoretical curve is depicted in Figure \ref{fig:fig4}b. The characteristic decay of the simulated curve is very well reproduced, the source of the deviation is the inexact treament of the nuclear quadrupole interaction.
%To our knowledge, no approximation has yet been derived valid for the case $\omega\approx A,B$.

%\vspace{0.5cm}

%\noindent{\emph{{Nuclear spin flip-flop and flip-flip interactions mediated by the defect spin}}
To explain the oscillation appearing in the second-order expansion in the nitrogen spin bath, we study the hyperfine interaction in clusters containing the defect and two nitrogen spins, one of which is a first neighbor. We note that, since the coupling strength of the direct magnetic dipolar coupling between two nuclear spins is typically of order kHz, the effect of this interaction is negligible on $T_2$ timescales considered. By comparing higher order decoherence contributions obtained with the full Hamiltonian and in the pseudo-secular approximation, see Figure~\ref{fig:fig2}a and Supporting Information, we conclude that the nonsecular $A_{xx}$, $A_{yy}$, and $A_{xy}$ hyperfine terms mediate an effective interaction between two, possibly distant nuclear spins, which then acts as a major source of decoherence.

To rigorously prove this, we employ the canonical Hamiltonian transformation on second-order clusters containing two nitrogen spins \cite{canonical-Sarma-2009}, and deduce effective nitrogen-nitrogen flip-flop $(\hat{I}_+^{(1)}\hat{I}_-^{(2)}+\hat{I}_-^{(1)}\hat{I}_+^{(2)})$ and flip-flip $(\hat{I}_+^{(1)}\hat{I}_+^{(2)}+\hat{I}_-^{(1)}\hat{I}_-^{(2)})$ interactions (for the detailed derivation see Supporting Information). The rate of these transitions is proportional to the product of the corresponding nonsecular hyperfine couplings of the nuclear spins and inversely proportional to the energy gap of qubit states ($|D-g_e\mu_BB_z|$). For the case of two first-neighbor nitrogen nuclear spins with $\mathcal{O}(100)$~MHz hyperfine couplings, we obtain sizable coupling even at low magnetic field.

Building on the results of the canonical transformation, we further derive the formula for the second-order ESEEM theory, see  Supporting Information. From this, we conclude that the sources of the observed oscillation are the clusters containing two first-neighbor nitrogens. The nuclear spin-nuclear spin $\ket{\uparrow\uparrow}\leftrightarrow\ket{\downarrow\downarrow}$ flip-flip transitions conditioned on the $\ket{-1}$ defect spin eigenstate introduce modulations of the coherence function with two slightly different frequencies $\nu_1\approx\nu_2$.
%$\nu_1=\frac{1}{4}\sqrt{\left(\Omega_{12}^{-1}\right)^2+\left(D_{12}^{-1}\right)^2}\approx33. 4$ MHz and $\nu_2=\frac{1}{4}\sqrt{\left(\Omega_{13}^{-1}\right)^2+\left(D_{13}^{-1}\right)^2}\approx33.4$ MHz.
The observed frequency emerges as the beat of these two frequencies; $\nu=\nu_1+\nu_2\approx67$~MHz. Consequently, the magnetic field-independent modulation of the coherence function is due to the first-neighbor nitrogen nuclear spin-nuclear spin flip-flip transitions enabled by strong Fermi contact hyperfine coupling induced by the in-plane localization of the spin density. The derived importance of first-neighbor nitrogen spins also explains the equality between $L$(t) obtained by completely neglecting these nuclei and in gCCE1, see Figure \ref{fig:fig4}b.

Finally, we simulated the coherence function with second-order ESEEM theory, where we see good agreement with simulated coherence functions in the modulation, see Figure~\ref{fig:fig4}e, f and  Supporting Information.

\subsubsection{Role of initial polarization of first-neighbor nitrogen spins}

After deducing the effect of first-neighbor nitrogens, we study the role of initial polarization of these spins, see Figure~\ref {fig:fig4}c. The experimentally achieved polarizations are 62\% at the excited state level-anticrossing (ESLAC) \cite{isotope-Gong-2024} and 30\% at the GSLAC \cite{isotope-Provost-2023}. Interestingly, in contrast to the common expectation that nuclear spin polarization enhances the coherence time, we obtain that $T_2$ is approximately independent of the initial population of the first neighbor nitrogens at $B_z = 50$~mT. 
We attribute this observation to the fact that the strongly coupled nitrogen spins do not dominate the coherence time in this magnetic field region.
%, but rather they modulate the decay function. Indeed, for increasing polarization, we observe increasing amplitude of the modulation, see Supporting Information.

%For few hundred mT magnetic fields, the energy gap between $\ket{\uparrow}$ and $\ket{\downarrow}$ nuclear spin eigenstates is of order MHz, therefore the hyperfine-induced $\ket{\uparrow}\leftrightarrow\ket{\downarrow}$ transitions, with mixing rates $\sim100$~MHz, cause swift relaxation of states, which leads to time evolution equivalent to the case of the thermal bath. \V{Hmmm...}

\subsection{The transition region (180-350~mT magnetic field interval)}

\begin{figure}[h!]
    \centering
    \includegraphics[width=\linewidth]{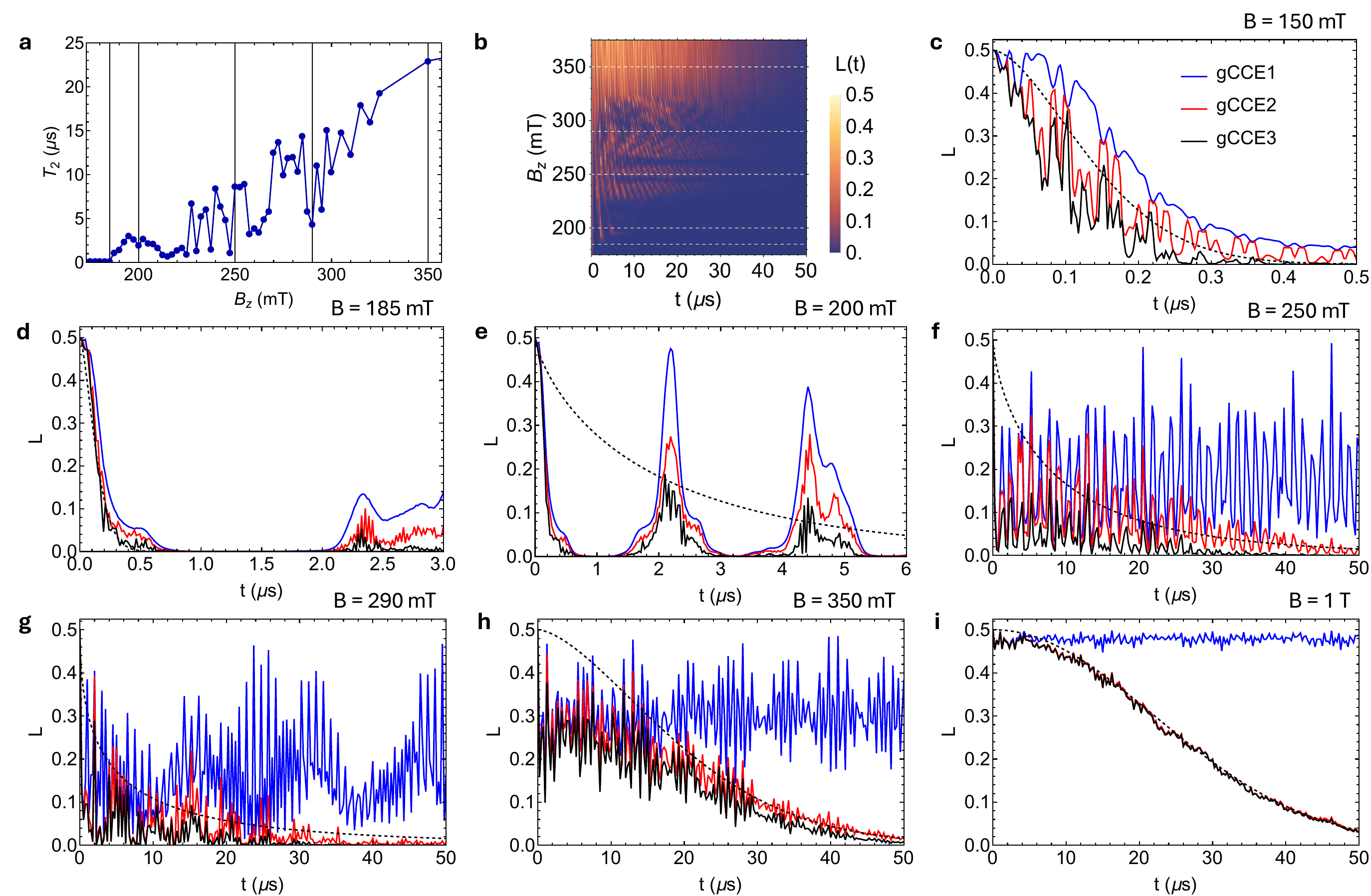}
    \caption{{\bf Coherence time in the transition region.} {\bf a} Coherence time as a function of magnetic field in the transition region. Lines indicate the sampled magnetic fields of the subsequent Figure. {\bf b} The coherence as a function of the spin-echo time and the external magnetic field. {\bf c, d, e, f, g, h, i} Coherence functions in different orders of gCCE. The magnetic field values indicated in the top-right corner of each subfigure were sampled from the transition region and the high-field regime. Dashed lines show the fitted coherence decays.}
    \label{fig:fig5}
\end{figure}

In the transition region, from 180~mT to 350~mT, three processes need to be taken into consideration; i) electron spin-mediated spin flip-flip processes of the strongly coupled nitrogen spins, ii) precession of boron nuclear spins above and below the defective hBN layer, and iii) nuclear spin-nuclear spin dipolar interactions giving rise to mutual nuclear spin flip-flops.   

Right after the GSLAC, the coherence time returns to the value of $\sim$200~ns, see Figure~\ref{fig:fig4}a. The coherence function exhibits analogous dynamics as before, i.e.\ in gCCE1 a Gaussian decay of the coherence function is observed, while noise originating from second- and third-order clusters introduces irregular oscillating patterns, see Figure~\ref{fig:fig5}c. Above 150~mT, the nuclear Zeeman splitting between the $\ket{\uparrow\uparrow}$ and the $\ket{\downarrow\downarrow}$ two nitrogen spin eigenstates exceeds the transition rate derived from the central spin-mediated nuclear spin flip-flip interaction, see  Supporting Information. As a consequence, the state mixing owing to this effective nuclear spin-nuclear spin coupling becomes suppressed.
%and the oscillation of $L(t)$ dissolves \V{Is this a better word?}.%saturates

%and the characteristic time of decay increases by an order of magnitude compared to the red curve in Figure \ref{fig:fig5}d.

Starting from $\sim185$~mT, the noninteracting bath gives rise to coherent beating on microsecond timescales, see Figure~\ref{fig:fig5}d, e.
%which is a consequence of the superposed modulation of different nuclear spins, as described in Equation \ref{eq:beating}. 
Higher-order spin correlations involving three nuclear spins induce additional decoherence. Based on previous studies of the spin bath model \cite{qmbtheory1-Ren-Bao}, the exponential decay obtained in Figs~\ref{fig:fig5}e, f, and g, suggests the increased importance of correlated noise of clusters with sufficiently large, discrete couplings to the defect spin. %For the exact treatment of effective interactions arising in gCCE3, the canonical Hamiltonian expansion should be derived up to third-order.

The quick collapse and partial, aperiodic revival of coherence (Figure~\ref{fig:fig5}f, g) has been observed for the silicon vacancy in silicon carbide \cite{4hsic-Yang-2014}, however, in contrast to the spin-3/2 $V_{Si}$ center, the V$_\mathrm{B}^-$ center is a spin-1 defect.
%In the pseudo-secular approximation, the decoherence contribution of third-order clusters is $\Tilde{L}(\mathrm{t})\equiv1$. Based on this and 
By comparing the magnitude of the nonsecular hyperfine terms characteristic of the shells surrounding the vacancy (Figure~\ref{fig:fig1}c) with the Zeeman splitting of different nuclear spin species, we detail a qualitative explanation of the observed dynamics independent of the spin quantum number of the defect spin. 

In the transition region, the increasing $|D-g_e\mu_BB_z|$ splitting of the electron spin states gradually reduces the strength of the effective coupling of the nitrogen nuclear spins mediated by the electron spin. In addition, the increasing nuclear Zeeman splitting of the nitrogen nuclear spins exceeds this effective coupling strength and thus suppresses the central spin-mediated processes. At the same time, the boron nuclear Zeeman splitting also suppresses the precession of the boron nuclear spins above and below the layer of the V$_\mathrm{B}^-$ center. As a result, the spin bath in the transition region becomes spatially separated; nuclear spin-flip processes occur at a few well-defined frequencies within the shells forming the inner, strongly interacting core, while the precession is suppressed in the outer, decoupled environment. The noise originating from the outer environment %are the ESEEM modulation of coherence 
is the homonuclear nuclear spin flip-flop induced by the dipole-dipole interaction. The frequency distribution of this noise, determined by the crystal structure, is a dense spectrum of closely spaced frequencies, which leads to Gaussian decay of coherence \cite{qmbtheory1-Ren-Bao}. We attribute the aperiodic beating of coherence to the correlated time evolution within the strongly coupled core, while decoherence on longer timescales is governed by the outer environment.
%We attribute the quick collapse of the coherence function to the Gaussian decay derived from ESEEM, and the partial revival to the correlated time evolution within the strongly coupled core.

Our model explains the upper limit of the transition region ($B_z\approx350$~mT), where the Zeeman splitting of $\ket{1/2}$ and $\ket{-1/2}$ boron eigenstates is %$350~\mathrm{mT}\cdot0.014~\mathrm{MHz~mT^{-1}}$
$\approx4.9$~MHz, which exceeds the nonsecular hyperfine constants of the boron spins in the shell closest to the vacancy, see Figure \ref{fig:fig1}b. For small-enough central cores, the system consisting of a few nuclear spins becomes coherent %even with respect to the central electronic spin,
and the weakly coupled environment becomes the leading noise source, see Figure \ref{fig:fig5}h.

\subsection{High-field regime}

Around $B_z\approx350$~mT the coherence time reaches the high-field limit. Above this value $T_2\approx31~\mu$s and the primary limiting factor of the coherence is the magnetic field-independent, dipole-dipole interaction-induced nuclear spin flip-flop interaction of homonuclear spin pairs, see Figure~\ref{fig:fig5}i for a typical coherence decay. Since the large Zeeman splitting of the $\ket{0}$ and $\ket{-1}$ defect eigenstates suppresses the nuclear spin-flipping processes excited by the second- and third-order effective interactions, the pseudo-secular approximation holds in the limit, see Figure~\ref{fig:fig2}a. In addition, the components of the first-order ESEEM modulation are transformed into a high-frequency, low-amplitude coherent oscillation. This is due to the nuclear Zeeman splitting suppressing the nuclear spin-flips caused by the pseudo-secular terms. Therefore, the modulation depth parameter in Equation~(\ref{eq:ESEEM1}) goes to $k_i\rightarrow0$. We note that the nuclear spin flip-flop induced by the dipole-dipole interaction primarily occurs between spin pairs located in the same layer of the material. In the case of van der Waals materials, the distance between the layers is typically 2-3 times larger than the lattice spacing \cite{2Drev-Seo-2024}, and the strength of the dipolar interaction decreases with $\sim d^{-3}$, where $d$ is the distance between the coupled spins.

In the  Supporting Information, we assess different transition channels and mixing rates, which highlight two important factors influencing the coherence time in the high-field limit. The spin quantum number of the nuclear spins in the bath determines the number of allowed transitions, which serve as individual channels of coherence loss. On the other hand, the mixing rate between nuclear spin states is proportional to the gyromagnetic ratios of the spin species. For the case of $^{10}$B and $^{11}$B isotopes, these have the opposite effect. Increase of the gyromagnetic ratio is of primary importance, resulting in reduced coherence time compared to h$^{10}$BN in the high field limit \cite{fpdecoh-Lee-2022}.

\section{Conclusion}

In this work, we provide a deep analysis of coherence dynamics for a wide range of external magnetic fields and other interaction parameters. We reveal and understand an intricate interplay of processes shaping nuclear magnetic noise and limiting the coherence time of the V$_\mathrm{B}^-$ center in h$^{11}$B$^{15}$N.  

To summarize our results, we observe the absence of coherence time enhancement at zero magnetic field even for large $E$ splittings. By scanning through a broad range of $E$ values, we identify the strong hyperfine coupling with the first-neighbor nitrogen nuclear spins as the primary source of this phenomenon. At low magnetic fields, the coherence-limiting magnetic noise is dominated by boron nuclear spins located in the layers above and below the V$_\mathrm{B}^-$ center's plane. In contrast, the strongly coupled in-plane nitrogen nuclear spins give rise to a characteristic envelope modulation of constant frequency. Surprisingly, we find that the initial polarization of the nearest nitrogen spins does not affect coherence time. Approaching the GSLAC, we demonstrate the emergence of electron spin-mediated nuclear spin flip-flop transitions involving the nearest nitrogen nuclear spins, which become significant as the energy gap between defect spin states decreases. Beyond the GSLAC, in the so-called transition region, three processes contribute to the decay and envelope modulation of the coherence function: electron spin-mediated nuclear spin flip-flop processes, the precession of boron nuclear spins above and below the V$_\mathrm{B}^-$ center's layer, and nuclear spin–nuclear spin transitions driven by mutual dipolar coupling. As the magnetic field increases, the first two processes vanish while the latter becomes dominant. Within the $180-350$~mT interval, the coherence time increases by two orders of magnitude and saturates at the magnetic-field-independent high-field limit of $T_2 \approx 31\mu$s, determined by nuclear spin–nuclear spin dipolar interactions. 

Understanding the dominating decoherence mechanisms enables us to define the optimal conditions of spin qubit operations and opens up the possibility of developing novel coherence protection protocols in hBN. As such, for sensing applications, we propose the utilization of the coherence properties of the V$_\mathrm{B}^-$ center in h$^{11}$B$^{15}$N at external magnetic field values within the transition region. This requires the application of only a moderate magnetic field ($B_z=180-350$~mT), yet yields one to two orders of magnitude enhancement to $\mathcal{O}(10~\mu$s$)$ in the coherence time compared to the low-field region ($\mathcal{O}(100$~ns$)$). 

\textit{Note added:} While preparing the manuscript, we became aware of a parallel study presenting complementary results. \cite{lee-decohfield-2025}

%\section{Methods}

\section*{Methodology}

To compute the hyperfine tensors from first principles, we performed density functional theory (DFT) calculations using a 972-atom hexagonal boron nitride supercell, adopting experimental lattice parameters. The core electrons were treated using the projector augmented-wave (PAW) method~\cite{Blochl1994}, while the valence electron wavefunctions were expanded in a plane-wave basis set with a cutoff energy of 450~eV. For the exchange-correlation functional, we employed the screened hybrid HSE06 functional~\cite{Heyd2003, Heyd2006} with $\alpha = 0.32$ mixing parameter, which offers a reliable description of the electronic structure in wide band gap materials such as hBN.

The V$_{\text{B}}^-$ geometry of the system in the ground state was fully relaxed until the residual atomic forces were below $5 \times 10^{-3}$~eV/\AA. Following structural relaxation, the hyperfine interaction tensors were calculated using the accurate method developed in Ref.~\cite{hypfine-Takacs-2024}. The hyperfine dataset used in the calculation is available in Ref.~\cite{Hyperfine-webpage}.

To theoretically describe the many-body system consisting of the experimentally controllable point defect and the surrounding atomic lattice, we employed the spin-bath model. In this model, the nuclear spins of the nuclei sitting at each lattice point serve as the noise-generating environment of the electronic spin forming the defect qubit.

%The effective Hamiltonian, containing single- and two-spin interactions present in the spin system, takes the following form:
%\begin{equation}
%\begin{split}
%    \hat{H}=&\hat{H}_{e-b}+\hat{H}_e+\hat{H}_b=\sum_i \mathbf{\hat{S}}^T\mathbf{A}^{(i)}\mathbf{\hat{I}}^{(i)}+\\ &D\left(\hat{S}_z^2-S(S+1)/3\right)+E\left(\hat{S}_x^2-\hat{S}_y^2\right)+g_e\mu_B\mathbf{B}^T\mathbf{\hat{S}}-\\
%    &\sum_ig_N^{(i)}\mu_N^{(i)} \mathbf{B}^T\mathbf{\hat{I}}^{(i)}+\sum_i\mathbf{\hat{I}}^{(i)}\mathbf{Q}^{(i)}\mathbf{\hat{I}}^{(i)}+\sum_{i<j}\mathbf{\hat{I}}^{T(i)}\mathbf{J}^{(ij)}\mathbf{\hat{I}}^{(j)},
%\end{split}
%\end{equation}
%where $\hat{H}_{e-b}$ is the interaction Hamiltonian describing hyperfine couplings between the V$_\mathrm{B}^-$ center and its environment. $\hat{H}_e$ and $\hat{H}_b$ are the separated electron spin and bath Hamiltonians, encoding the zero-field splitting of the V$_\mathrm{B}^-$ center, the Zeeman-interaction, the nuclear quadrupole splitting for each nucleus, and the magnetic dipolar interaction of nuclear spins, respectively. Note that accurate hyperfine tensors for the $^{15}$N and $^{11}$B nuclear spins were derived from first-principles calculations, avoiding finite-size effects~\cite{hypfine-Takacs-2024}.

To numerically study the Hahn-echo coherence time of bulk V$_\mathrm{B}^-$ centers in hexagonal boron nitride, we implemented the third-order generalized cluster correlation expansion method (gCCE3)~\cite{yang-quantum-2008, coherence-avoided-cr-Onizhuk-2021,decohVB-Haykal-2022}. Coherence time calculations were performed considering a monoisotopic nuclear spin bath containing $^{15}$N nitrogen and $^{11}$B boron isotopes. The simulations included a few hundred first-order, several thousand second-order, and up to twenty thousand third-order subsystems. The extensive calculations ensuring numerical convergence are detailed in Supporting Information.

%To uncover the decoherence processes owing to hyperfine interaction-driven transitions between electronic spin eigenstates, we used the pseudo-secular approximation of hyperfine couplings. In this approximation the Hamiltonian terms containing the $\hat{S}_x$ and $\hat{S}_y$ spin-flipping operators are neglected. The approximation holds when the energy gap between different eigenstates is large enough to suppress transitions and no higher-order effective interactions emerge in the system.

\RaggedRight
\medskip
\textbf{Supporting Information} \par %Please delete the Suppporting Information statement if it is not applicable. Please supply Supporting Information in another file. Supporting information should not be provided in .tex format
Supporting Information is available from the Wiley Online Library or from the author.

% Acknowledgements
\medskip
\textbf{Acknowledgements} \par %delete if not applicable))
This research was supported by the National Research, Development, and Innovation Office of Hungary within the Quantum Information National Laboratory of Hungary (Grant No. 2022-2.1.1-NL-2022-00004) and within grant FK 145395.
This project is funded by the European Commission within Horizon Europe projects (Grant Nos.\ 101156088 and 101129663).
%We acknowledge support from the Knut and Alice Wallenberg Foundation through WBSQD2 project (Grant No.\ 2018.0071). 
The computations were enabled by resources provided by the 7 Academic Infrastructure for Supercomputing in Sweden (NAISS) and the Swedish National Infrastructure for Computing (SNIC) at the National Supercomputer Centre (NSC) partially funded by the Swedish Research Council through grant agreements no. 2022-06725 and no. 2018-05973. We acknowledge KIF\"U for awarding us access to computational resources in Hungary.

\medskip
\textbf{Data Availability Statement}\par
The data that support the findings of this study are available from the authors upon reasonable request.

\medskip
\textbf{Code Availability Statement}\par

The codes associated with this manuscript are available from the corresponding author upon reasonable request.

\medskip
\textbf{Author contributions}\par

A.T.\ carried out the numerical simulations. A.T.\ and V.I.\ wrote the manuscript. The work was supervised by V.I.

\medskip
\textbf{Competing Interest}\par

The authors declare no competing interests.

% References
\medskip

% Use the following code if you wish to generate your bibliography with BibTeX;
% replace the string "MSP-template" below with the name(s) of
% the BibTeX data base(s) you want to use.
% The resulting bibliography-output (the content of the .bbl file)
% must be pasted back into this file before submission.
% Please also include your BibTeX data base file(s) in your submission
% so that we can re-run BibTeX if necessary.
%

%\bibliographystyle{MSP}
%\bibliography{MSP-template}
%\bibliography{references}

\begin{thebibliography}{10}
\providecommand{\url}[1]{\texttt{#1}}
\providecommand{\urlprefix}{URL }

\bibitem{degen-quantum-2017}
C.~L. Degen, F.~Reinhard, P.~Cappellaro,
\newblock \emph{Rev. Mod. Phys.} \textbf{2017}, \emph{89} 035002.

\bibitem{aslam-quantum-2023}
N.~Aslam, H.~Zhou, E.~K. Urbach, M.~J. Turner, R.~L. Walsworth, M.~D. Lukin,
  H.~Park,
\newblock \emph{Nature Reviews Physics} \textbf{2023}, 1--13, publisher: Nature
  Publishing Group.

\bibitem{du-single-molecule-2024}
J.~Du, F.~Shi, X.~Kong, F.~Jelezko, J.~Wrachtrup,
\newblock \emph{Reviews of Modern Physics} \textbf{2024}, \emph{96}, 2 025001,
  publisher: American Physical Society.

\bibitem{doherty-nitrogen-vacancy-2013}
M.~W. Doherty, N.~B. Manson, P.~Delaney, F.~Jelezko, J.~Wrachtrup, L.~C.~L.
  Hollenberg,
\newblock \emph{Physics Reports} \textbf{2013}, \emph{528}, 1 1.

\bibitem{schirhagl-nitrogen-vacancy-2014}
R.~Schirhagl, K.~Chang, M.~Loretz, C.~L. Degen,
\newblock \emph{Annual Review of Physical Chemistry} \textbf{2014}, \emph{65},
  1 83, \_eprint: https://doi.org/10.1146/annurev-physchem-040513-103659.

\bibitem{dolde-electric-field-2011}
F.~Dolde, H.~Fedder, M.~W. Doherty, T.~Nöbauer, F.~Rempp, G.~Balasubramanian,
  T.~Wolf, F.~Reinhard, L.~C.~L. Hollenberg, F.~Jelezko, J.~Wrachtrup,
\newblock \emph{Nature Physics} \textbf{2011}, \emph{7}, 6 459,
  bandiera\_abtest: a Cg\_type: Nature Research Journals Number: 6
  Primary\_atype: Research Publisher: Nature Publishing Group.

\bibitem{boss-quantum-2017}
J.~M. Boss, K.~S. Cujia, J.~Zopes, C.~L. Degen,
\newblock \emph{Science} \textbf{2017}, \emph{356}, 6340 837, publisher:
  American Association for the Advancement of Science Section: Reports.

\bibitem{rendler-optical-2017}
T.~Rendler, J.~Neburkova, O.~Zemek, J.~Kotek, A.~Zappe, Z.~Chu, P.~Cigler,
  J.~Wrachtrup,
\newblock \emph{Nature Communications} \textbf{2017}, \emph{8}, 1 14701,
  bandiera\_abtest: a Cc\_license\_type: cc\_by Cg\_type: Nature Research
  Journals Number: 1 Primary\_atype: Research Publisher: Nature Publishing
  Group Subject\_term: Biosensors;Magnetic materials;Nanoparticles
  Subject\_term\_id: biosensors;magnetic-materials;nanoparticles.

\bibitem{hache-nanoscale-2025}
T.~Hache, A.~Anshu, T.~Shalomayeva, G.~Richter, R.~Stöhr, K.~Kern,
  J.~Wrachtrup, A.~Singha,
\newblock \emph{Nano Letters} \textbf{2025}, \emph{25}, 5 1917, publisher:
  American Chemical Society.

\bibitem{petrini-nanodiamondquantum-2022}
G.~Petrini, G.~Tomagra, E.~Bernardi, E.~Moreva, P.~Traina, A.~Marcantoni,
  F.~Picollo, K.~Kvaková, P.~Cígler, I.~P. Degiovanni, V.~Carabelli,
  M.~Genovese,
\newblock \emph{Advanced Science} \textbf{2022}, \emph{9}, 28 2202014,
  \_eprint: https://onlinelibrary.wiley.com/doi/pdf/10.1002/advs.202202014.

\bibitem{sangtawesin-origins-2019}
S.~Sangtawesin, B.~L. Dwyer, S.~Srinivasan, J.~J. Allred, L.~V. Rodgers,
  K.~De~Greve, A.~Stacey, N.~Dontschuk, K.~M. O‚ÄôDonnell, D.~Hu, D.~A. Evans,
  C.~Jaye, D.~A. Fischer, M.~L. Markham, D.~J. Twitchen, H.~Park, M.~D. Lukin,
  N.~P. de~Leon,
\newblock \emph{Physical Review X} \textbf{2019}, \emph{9}, 3 031052,
  publisher: American Physical Society.

\bibitem{bluvstein-identifying-2019}
D.~Bluvstein, Z.~Zhang, A.~C.~B. Jayich,
\newblock \emph{Phys. Rev. Lett.} \textbf{2019}, \emph{122} 076101.

\bibitem{dwyer-probing-2022}
B.~L. Dwyer, L.~V. Rodgers, E.~K. Urbach, D.~Bluvstein, S.~Sangtawesin,
  H.~Zhou, Y.~Nassab, M.~Fitzpatrick, Z.~Yuan, K.~De~Greve, E.~L. Peterson,
  H.~Knowles, T.~Sumarac, J.-P. Chou, A.~Gali, V.~Dobrovitski, M.~D. Lukin,
  N.~P. de~Leon,
\newblock \emph{PRX Quantum} \textbf{2022}, \emph{3}, 4 040328, publisher:
  American Physical Society.

\bibitem{pershin-shallowNV-2025}
A.~Pershin, A.~Tárkányi, V.~Verkhovlyuk, V.~Ivády, A.~Gali,
\newblock A coherence-protection scheme for quantum sensors based on
  ultra-shallow single nitrogen-vacancy centers in diamond, {arXiv:2501.00180} \textbf{2025},
\newblock \urlprefix\url{https://arxiv.org/abs/2501.00180}.

\bibitem{tetienne-quantum-2021}
J.-P. Tetienne,
\newblock \emph{Nature Physics} \textbf{2021}, 1--2, bandiera\_abtest: a
  Cg\_type: Nature Research Journals Primary\_atype: News \& Views Publisher:
  Nature Publishing Group Subject\_term: Qubits;Two-dimensional materials
  Subject\_term\_id: qubits;two-dimensional-materials.

\bibitem{hassan-2d-2023}
J.~Z. Hassan, A.~Raza, Z.~U.~D. Babar, U.~Qumar, N.~T. Kaner, A.~Cassinese,
\newblock \emph{Journal of Materials Chemistry A} \textbf{2023}, \emph{11}, 12
  6016, publisher: The Royal Society of Chemistry.

\bibitem{vaidya-quantum-2023}
S.~Vaidya, X.~Gao, S.~Dikshit, I.~Aharonovich, T.~Li,
\newblock \emph{Advances in Physics: X} \textbf{2023}, \emph{8}, 1 2206049,
  publisher: Taylor \& Francis \_eprint:
  https://doi.org/10.1080/23746149.2023.2206049.

\bibitem{gottscholl-spin-2021}
A.~Gottscholl, M.~Diez, V.~Soltamov, C.~Kasper, D.~Krauße, A.~Sperlich,
  M.~Kianinia, C.~Bradac, I.~Aharonovich, V.~Dyakonov,
\newblock \emph{Nature Communications} \textbf{2021}, \emph{12}, 1 4480,
  bandiera\_abtest: a Cc\_license\_type: cc\_by Cg\_type: Nature Research
  Journals Number: 1 Primary\_atype: Research Publisher: Nature Publishing
  Group Subject\_term: Electronic properties and materials;Qubits
  Subject\_term\_id: electronic-properties-and-materials;qubits.

\bibitem{healey-quantum-2022}
A.~J. Healey, S.~C. Scholten, T.~Yang, J.~A. Scott, G.~J. Abrahams, I.~O.
  Robertson, X.~F. Hou, Y.~F. Guo, S.~Rahman, Y.~Lu, M.~Kianinia,
  I.~Aharonovich, J.-P. Tetienne,
\newblock \emph{Nature Physics} \textbf{2022}, 1--5, publisher: Nature
  Publishing Group.

\bibitem{kumar-magnetic-2022}
P.~Kumar, F.~Fabre, A.~Durand, T.~Clua-Provost, J.~Li, J.~Edgar,
  N.~Rougemaille, J.~Coraux, X.~Marie, P.~Renucci, C.~Robert, I.~Robert-Philip,
  B.~Gil, G.~Cassabois, A.~Finco, V.~Jacques,
\newblock \emph{Physical Review Applied} \textbf{2022}, \emph{18}, 6 L061002,
  publisher: American Physical Society.

\bibitem{gottscholl-initialization-2020}
A.~Gottscholl, M.~Kianinia, V.~Soltamov, S.~Orlinskii, G.~Mamin, C.~Bradac,
  C.~Kasper, K.~Krambrock, A.~Sperlich, M.~Toth, I.~Aharonovich, V.~Dyakonov,
\newblock \emph{Nature Materials} \textbf{2020}, \emph{19}, 5 540, number: 5
  Publisher: Nature Publishing Group.

\bibitem{ivady-ab-2020}
V.~Ivády, G.~Barcza, G.~Thiering, S.~Li, H.~Hamdi, J.-P. Chou, Ö.~Legeza,
  A.~Gali,
\newblock \emph{npj Computational Materials} \textbf{2020}, \emph{6}, 1 1,
  number: 1 Publisher: Nature Publishing Group.

\bibitem{lyu-strain-2022}
X.~Lyu, Q.~Tan, L.~Wu, C.~Zhang, Z.~Zhang, Z.~Mu, J.~Zúñiga-Pérez, H.~Cai,
  W.~Gao,
\newblock \emph{Nano Letters} \textbf{2022}, \emph{22}, 16 6553, publisher:
  American Chemical Society.

\bibitem{yang-spin-2022}
T.~Yang, N.~Mendelson, C.~Li, A.~Gottscholl, J.~Scott, M.~Kianinia,
  V.~Dyakonov, M.~Toth, I.~Aharonovich,
\newblock \emph{Nanoscale} \textbf{2022}, \emph{14}, 13 5239, publisher: The
  Royal Society of Chemistry.

\bibitem{udvarhelyi-planar-2023}
P.~Udvarhelyi, T.~Clua-Provost, A.~Durand, J.~Li, J.~H. Edgar, B.~Gil,
  G.~Cassabois, V.~Jacques, A.~Gali,
\newblock \emph{npj Computational Materials} \textbf{2023}, \emph{9}, 1 1,
  publisher: Nature Publishing Group.

\bibitem{gao-quantum-2023}
X.~Gao, S.~Vaidya, P.~Ju, S.~Dikshit, K.~Shen, Y.~P. Chen, T.~Li,
\newblock \emph{ACS Photonics} \textbf{2023}, \emph{10}, 8 2894, publisher:
  American Chemical Society.

\bibitem{durand-optically-2023}
A.~Durand, T.~Clua-Provost, F.~Fabre, P.~Kumar, J.~Li, J.~Edgar, P.~Udvarhelyi,
  A.~Gali, X.~Marie, C.~Robert, J.~Gérard, B.~Gil, G.~Cassabois, V.~Jacques
  \emph{131}, 11 116902, publisher: American Physical Society.

\bibitem{robertson-detection-2023}
I.~O. Robertson, S.~C. Scholten, P.~Singh, A.~J. Healey, F.~Meneses,
  P.~Reineck, H.~Abe, T.~Ohshima, M.~Kianinia, I.~Aharonovich, J.-P. Tetienne
  \emph{17}, 14 13408, publisher: American Chemical Society.

\bibitem{clua-provost-impact-2024}
T.~Clua-Provost, A.~Durand, J.~Fraunié, C.~Robert, X.~Marie, J.~Li, J.~H.
  Edgar, B.~Gil, J.-M. Gérard, G.~Cassabois, V.~Jacques \emph{24}, 41 12915,
  publisher: American Chemical Society.

\bibitem{fraunie-charge-2025}
J.~Fraunié, T.~Clua-Provost, S.~Roux, Z.~Mu, A.~Delpoux, G.~Seine, D.~Lagarde,
  K.~Watanabe, T.~Taniguchi, X.~Marie, T.~Poirier, J.~H. Edgar, J.~Grisolia,
  B.~Lassagne, A.~Claverie, V.~Jacques, C.~Robert,
\newblock \emph{Nano Letters} \textbf{2025}, \emph{25}, 14 5836, publisher:
  American Chemical Society.

\bibitem{huang-wide-2022}
M.~Huang, J.~Zhou, D.~Chen, H.~Lu, N.~J. {McLaughlin}, S.~Li, M.~Alghamdi,
  D.~Djugba, J.~Shi, H.~Wang, C.~R. Du \emph{13}, 1 5369, publisher: Nature
  Publishing Group.

\bibitem{VBmeas-liu2022}
W.~Liu, V.~Iv{\'a}dy, Z.~P. Li, et~al.,
\newblock \emph{Nature Communications} \textbf{2022}, \emph{13} 5713.

\bibitem{VBmeas-Gottscholl-2021}
A.~Gottscholl, M.~Diez, V.~Soltamov, S.~Kasper, P.~Sperlich, K.~Krau{\ss}e,
  A.~K{\"u}rten, P.~G. Caspar, V.~A. Nadolinny, G.~Wagner, M.~H. Jakob,
  J.~Meijer, G.~G. Borghs, V.~Dyakonov,
\newblock \emph{Science Advances} \textbf{2021}, \emph{7}, 19 eabf3630.

\bibitem{extending-Rizzato2023}
R.~Rizzato, M.~Schalk, S.~Mohr, et~al.,
\newblock \emph{Nature Communications} \textbf{2023}, \emph{14} 5089.

\bibitem{isotope-Gong-2024}
R.~Gong, X.~Du, E.~Janzen, et~al.,
\newblock \emph{Nature Communications} \textbf{2024}, \emph{15} 104.

\bibitem{decohVB-Haykal-2022}
A.~Haykal, R.~Tanos, N.~Minotto, et~al.,
\newblock \emph{Nature Communications} \textbf{2022}, \emph{13} 4347.

\bibitem{cohprot-Ramsay2023}
A.~J. Ramsay, R.~Hekmati, C.~J. Patrickson, et~al.,
\newblock \emph{Nature Communications} \textbf{2023}, \emph{14} 461.

\bibitem{VBmeas3T-dyakonov-2022}
F.~F. Murzakhanov, G.~V. Mamin, S.~B. Orlinskii, U.~Gerstmann, W.~G. Schmidt,
  T.~Biktagirov, I.~Aharonovich, A.~Gottscholl, A.~Sperlich, V.~Dyakonov, V.~A.
  Soltamov,
\newblock \emph{Nano Letters} \textbf{2022}, \emph{22}, 7 2718, pMID: 35357842.

\bibitem{yang-quantum-2008}
W.~Yang, R.-B. Liu \emph{78}, 8 085315, publisher: American Physical Society.

\bibitem{zhao-decoherence-2012}
N.~Zhao, S.-W. Ho, R.-B. Liu \emph{85}, 11 115303, publisher: American Physical
  Society.

\bibitem{seo-quantum-2016}
H.~Seo, A.~L. Falk, P.~V. Klimov, K.~C. Miao, G.~Galli, D.~D. Awschalom
  \emph{7}, 1 1, number: 1 Publisher: Nature Publishing Group.

\bibitem{onizhuk-probing-2021}
M.~Onizhuk, K.~C. Miao, J.~P. Blanton, H.~Ma, C.~P. Anderson, A.~Bourassa,
  D.~D. Awschalom, G.~Galli \emph{2}, 1 010311, publisher: American Physical
  Society.

\bibitem{yang-longitudinal-2020}
Z.-S. Yang, Y.-X. Wang, M.-J. Tao, W.~Yang, M.~Zhang, Q.~Ai, F.-G. Deng
  \emph{413} 168063.

\bibitem{hypfine-Takacs-2024}
I.~Takács, V.~Ivády,
\newblock \emph{Communications Physics} \textbf{2024}, \emph{7} 178.

\bibitem{Hyperfine-webpage}
The hyperfine dataset, including 13000 entries, can be found at,
\newblock \url{https://ivadygroup.elte.hu/hyperfine/}.

\bibitem{jamonneau-competition-2016}
P.~Jamonneau, M.~Lesik, J.~P. Tetienne, I.~Alvizu, L.~Mayer, A.~Dréau,
  S.~Kosen, J.-F. Roch, S.~Pezzagna, J.~Meijer, T.~Teraji, Y.~Kubo, P.~Bertet,
  J.~R. Maze, V.~Jacques \emph{93}, 2 024305.

\bibitem{fpdecoh-Lee-2022}
J.~Lee, H.~Park, H.~Seo,
\newblock \emph{npj 2D Materials and Applications} \textbf{2022}, \emph{6} 60.

\bibitem{coherence-avoided-cr-Onizhuk-2021}
M.~Onizhuk, K.~C. Miao, J.~P. Blanton, H.~Ma, C.~P. Anderson, A.~Bourassa,
  D.~D. Awschalom, G.~Galli,
\newblock \emph{PRX Quantum} \textbf{2021}, \emph{2} 010311.

\bibitem{4hsic-Yang-2014}
L.-P. Yang, C.~Burk, M.~Widmann, S.-Y. Lee, J.~Wrachtrup, N.~Zhao,
\newblock \emph{Phys. Rev. B} \textbf{2014}, \emph{90} 241203.

\bibitem{seo-first-principles-2024}
H.~Seo, V.~Ivády, Y.~Ping \emph{125}, 14 140501.

\bibitem{quadeseem-SHUBIN-1983}
A.~Shubin, S.~Dikanov,
\newblock \emph{Journal of Magnetic Resonance (1969)} \textbf{1983}, \emph{52},
  1 1.

\bibitem{sic-divacancy-Seo}
H.~Seo, A.~Falk, P.~Klimov, K.~Miao, G.~Galli, D.~Awschalom,
\newblock \emph{Nature Communications} \textbf{2016}, \emph{7}.

\bibitem{canonical-Sarma-2009}
L.~Cywi\ifmmode~\acute{n}\else \'{n}\fi{}ski, W.~M. Witzel, S.~Das~Sarma,
\newblock \emph{Phys. Rev. B} \textbf{2009}, \emph{79} 245314.

\bibitem{isotope-Provost-2023}
T.~Clua-Provost, A.~Durand, Z.~Mu, T.~Rastoin, J.~Frauni\'e, E.~Janzen,
  H.~Schutte, J.~H. Edgar, G.~Seine, A.~Claverie, X.~Marie, C.~Robert, B.~Gil,
  G.~Cassabois, V.~Jacques,
\newblock \emph{Phys. Rev. Lett.} \textbf{2023}, \emph{131} 126901.

\bibitem{qmbtheory1-Ren-Bao}
W.~Yang, W.-L. Ma, R.-B. Liu,
\newblock \emph{Rep. Prog. Phys.} \textbf{2016}, \emph{80} 016001.

\bibitem{2Drev-Seo-2024}
H.~Seo, V.~Ivády, Y.~Ping,
\newblock \emph{Applied Physics Letters} \textbf{2024}, \emph{125}, 14 140501.

\bibitem{lee-decohfield-2025}
J.~Lee, H.~Kim, H.~Park, H.~Seo,
\newblock Magnetic-field dependent vb- spin decoherence in hexagonal boron
  nitrides: A first-principles study, {arXiv:2505.03306} \textbf{2025},
\newblock \urlprefix\url{https://arxiv.org/abs/2505.03306}.

\bibitem{Blochl1994}
P.~E. Blöchl,
\newblock \emph{Phys. Rev. B} \textbf{1994}, \emph{50} 17953.

\bibitem{Heyd2003}
J.~Heyd, G.~E. Scuseria, M.~Ernzerhof,
\newblock \emph{J. Chem. Phys.} \textbf{2003}, \emph{118} 8207.

\bibitem{Heyd2006}
J.~Heyd, G.~E. Scuseria, M.~Ernzerhof,
\newblock \emph{J. Chem. Phys.} \textbf{2006}, \emph{124} 219906.

\end{thebibliography}

% Figure/tables and captions
% Permission statements are required for all Figure reproduced or adapted from previously published articles/sources. Please also ensure that all necessary permissions to reproduce images have been received
% Please remove these statements for original Figure

% Table of contents entry should be 50 - 60 words long
% Image should be 55 mm broad and 50 mm high or 110 mm broad and 20 mm high

%\begin{figure}[h!]
%\textbf{Table of Contents}\\
%\medskip
%  \includegraphics{toc-image.png}
%  \medskip
%  \caption*{ToC Entry}
%\end{figure}

\end{document}